# A Review of Quantum Scientific Computing Algorithms for Engineering Problems

Osama Muhammad Raisuddin[1]*, Suvranu De[2]

*raisuo2@rpi.edu

[1] *Mechanical, Aerospace, and Nuclear Engineering, Rensselaer Polytechnic Institute, 110 8th St. Troy, NY, 12180, USA*

[2] *Florida Agricultural and Mechanical University-Florida State University College of Engineering, 2525 Pottsdamer St., Tallahassee, FL 32310, USA*

## Abstract

Quantum computing, leveraging quantum phenomena like superposition and entanglement, is emerging as a transformative force in computing technology, promising unparalleled computational speed and efficiency crucial for engineering applications. This advancement presents both opportunities and challenges, requiring engineers to familiarize themselves with quantum principles, applications, and complexities. This paper systematically explores the foundational concepts of quantum mechanics and their implications for computational advancements, emphasizing the superiority of quantum algorithms in solving engineering problems. It identifies areas where gate-based quantum computing has the potential to outperform classical methods despite facing scalability and coherence issues. By offering clear examples with minimal reliance on in-depth quantum physics or hardware specifics, the aim is to make quantum computing accessible to engineers, addressing the steep learning curve and fostering its practical adoption for complex problem-solving and technological advancement as quantum hardware becomes more robust and reliable.

# 1. Introduction

Quantum computing is an emerging area of research with great potential for engineering and scientific computing applications. Quantum mechanical effects, including superposition, entanglement, and interference, enable quantum computers to achieve exponential speedups over classical computers for certain classes of problems [1]. The advantages can potentially be realized in computation time, resources, and energy consumption. All classical computations can be performed on quantum computers using reversible versions of logic gates. However, this approach is inefficient and requires large overheads. To take advantage of quantum computers, new algorithms have been developed and are an active area of research. Due to significant differences with classical computation, the engineering community has limited exposure to quantum computing and quantum algorithms. In this paper, we provide an overview of quantum algorithms for engineering with supporting codes to demonstrate the fundamental ideas.

Quantum computing was envisaged as a tool for simulating quantum systems [2] which can be exponentially expensive to simulate classically. Quantum algorithms have been developed for some classes of problems, including factorization [3] and the solution of systems of linear systems [4], [5] with exponential speedups over their classical counterparts. Classical computing power has steadily increased over the past few decades in accordance with Moore's law; the number of transistors in an integrated circuit doubles every two years. One factor driving this growth is the miniaturization of transistors to a few nanometers. However, further miniaturization is implausible due to the quantum regime superseding the deterministic regime in which transistors operate. Quantum computing is a viable candidate for tackling computations beyond the end of Moore's law for select problems.

The development of quantum computer hardware is a dynamic research area exploring various potential systems, including photons, trapped ions, topological conductors, or superconducting circuits like transmon, fluxonium, or SQUIDs [1]. Despite the diversity of physical implementations, the development of computational algorithms can proceed through the abstraction of these systems into a computational model of quantum computing. The field predominantly adopts two approaches: gate-based quantum computing and adiabatic quantum computing, each representing a distinct methodology for quantum computation [6].

Gate-based quantum computing is the most prominent computing model with a provable advantage over classical computing for various problems. The gate-based quantum computing model is grounded in complex linear algebra and probability. The state of a quantum computer is described using quantum bits or qubits, defined as a unit vector of probability amplitudes and operations. Operations of quantum computers, quantum gates and measurements, are defined as unitary and projection matrices, respectively. The input to a quantum algorithm can be provided as a combination of the initial state of the qubits and a collection of (possibly parametrized) gates, which form a quantum circuit. The output state of the quantum computer is read out by performing measurements of the qubits to obtain a distribution of results. Universal quantum gate sets are similar to universal logic gate sets for classical computing [1]. Therefore, a quantum algorithm can be defined using any convenient set of gates, which can then be converted to the universal gate set of an underlying quantum architecture. The gate-based architecture allows abstraction of the computational model from the underlying hardware.

In this paper, we introduce the concepts and ideas of gate-based quantum computing and review available quantum algorithms for engineering problems. The complexity analysis of such algorithms differs from classical computers. The complexity of quantum algorithms is typically defined as a combination of the number of qubits, quantum gates, quantum oracle queries, and the success probability of measuring some qubits in a desired state. We also demonstrate key ideas and algorithms with code for quantum computer simulators.

Quantum computers are susceptible to noise, imperfections in fabrication, and errors in control hardware. Error correction and noise mitigation is crucial to the operations of quantum computers [7]. Creating logical qubits from noisy physical qubits is a multifaceted and active area of research. Recent work has demonstrated the scalability of error correction for bit flip, phase, and a combination of both through repetition [8] and surface codes [9] for superconducting qubits. Quantum error correction and mitigation is an active area of research [10], [11], [12], [13] and will not be discussed in this paper.

In Section 2 we introduce the elements of gate-based quantum computing. In Section 3 we provide an overview of the programming stack for quantum computing and prevalent implementations and libraries for various components of the stack. In Section 4 we introduce quantum subroutines used as building blocks of quantum algorithms. We then demonstrate the usage of these subroutines to build algorithms for scientific and engineering computation and their applications in Section 5. Section 5 first introduces the quantum Hamiltonian simulation problem and the quantum linear system problem in Sections 5.1 and 5.2, respectively, with code implementations. Later in Sections 5.3 and 5.4 algorithms for ordinary and partial differential equations are discussed, respectively. We then introduce variational quantum algorithms and an overview of notable variational quantum algorithms in Section 6. In Section 7, we discuss some challenges and open questions in quantum computing.

## 2. Gate-based Quantum Computing

Gate-based quantum computing is based on a system of interacting quantum objects, which is described using complex linear algebra. In this section we abstract away details of quantum information and the underlying physics of a quantum computer and present the computational model for quantum computing in Dirac or 'bra-ket' notation. This is without any loss of generality since all gate-based quantum computers share the same computational model. Implementation of quantum algorithms does not require knowledge beyond these fundamental elements. However, we occasionally provide brief references to the underlying physics to develop insights.

In Section 2.1 we introduce the quantum counterpart of classical computing bits, qubits, and show how their statevector (state vector) exists in exponentially larger Hilbert spaces in Section 2.2. We then provide examples of techniques to encode data into registers of qubits in Section 2.3. In Section 2.4 we introduce quantum gates, the building blocks of quantum computers, which operate on qubits. We introduce some elementary gates and compare their properties with classical logic gates. In Section 2.5 we introduce the concept of measuring quantum states or qubits, and its corresponding mathematical operation. We combine these basic ideas in Section 2.6 to create quantum circuits to demonstrate the properties of superposition and entanglement in Sections 2.7 and 2.8. In Section 2.9 we discuss the idea of reversible computation and how classically irreversible gates, e.g., AND or OR, can be implemented using quantum gates to demonstrate that all classical computation can be performed on quantum computers, albeit with high overheads. In Section 2.10 we discuss examples of quantum oracles, which are black-box methods to access information in quantum algorithms. We conclude Section 2 by discussing the limitations of quantum computing in Section 2.11.

## 2.1.    Qubits

In this section, we introduce the mathematical model of a qubit. In Section 2.2 we will consider a collection of qubits as a quantum register.

**Definition 1:** A qubit or quantum bit $|\psi\rangle$ is a system whose quantum state $|\psi\rangle \in \mathcal{H} \cong \mathbb{C}^2$ can be described by a superposition of two orthogonal eigenstates labeled as $|0\rangle$ and $|1\rangle$, i.e., $|\psi\rangle = \alpha|0\rangle + \beta|1\rangle$ where $|\alpha|^2 + |\beta|^2 = 1$ and $\alpha, \beta \in \mathbb{C}$.

A 'qubit', or a quantum bit, is the quantum analog of a classical bit [1] and is the elementary unit of information in quantum computing. A qubit is a two-dimensional quantum system with dimensions typically labeled '0' and '1', in gate-based quantum computing, corresponding to the binary states of a classical bit. The state of a qubit exists in the Hilbert space $\mathcal{H}: \mathbb{C}^2$ and is represented in 'bra-ket' or Dirac notation as:

$$|\psi\rangle = \alpha|0\rangle + \beta|1\rangle \tag{1}$$

where $|\psi\rangle \in \mathbb{C}^2$ denotes the quantum state of the qubit with complex probability amplitudes $\alpha, \beta \in \mathbb{C}$ that are normalized according to the Born rule [14], [15] as

$$|\alpha|^2 + |\beta|^2 = 1 \tag{2}$$

$|0\rangle$ and $|1\rangle$ are orthonormal basis states in the $0 - 1$ basis, also referred to as the $Z$ basis, corresponding to the basis vectors $|0\rangle = \begin{pmatrix} 1 \\ 0 \end{pmatrix}$ and $|1\rangle = \begin{pmatrix} 0 \\ 1 \end{pmatrix}$. $|\alpha|^2$ and $|\beta|^2$ represent the probabilities of measuring the qubit as 0 and 1, respectively. Measurement is discussed in detail in Section 2.5. Probability amplitudes differ from probabilities since $\alpha, \beta \in \mathbb{C}$ instead of $\mathbb{R}$, and can have phases associated with them. The existence of phases in probability amplitudes allows quantum states to interfere constructively or destructively [16].

Once a measurement is performed on qubit $|\psi\rangle$ and it is measured as 0 or 1, the state $|\psi\rangle$

"collapses" to $|0\rangle$ or $|1\rangle$, destroying the superposition. Further measurements (without any other subsequent operations) in the same basis ($|0\rangle$ and $|1\rangle$) will repeatedly yield the same classical state that was originally measured. As an example, physically, a qubit can be represented by the discrete energy levels (ground state or first excited state) of a closed shell atom (a boson), and measurements correspond to measurement of the energy level of the atom.

The column vector $|\psi\rangle$ is a "ket". The corresponding "bra" $\langle\psi|$ is the Hermitian transpose, or the conjugate transpose of the ket, which is the functional that takes a ket to the space of real numbers. As an example, $\langle 0| = (|0\rangle)^{\dagger}$. The Hermitian transpose is represented using the $\dagger$ "dagger" symbol as $\langle\psi| = (|\psi\rangle)^{\dagger}$. This notation is used throughout quantum computing literature and is referred to as "bra-ket" or Dirac notation. Inner and outer products in bra-ket notation are represented as $\langle\cdot\,|\,\cdot\rangle$ and $|\cdot\rangle\langle\cdot|$, respectively.

In the Dirac notation, the label used in a bra or ket can correspond to physical properties or the information encoded in the qubit. The labels $|1\rangle, |-1\rangle$ correspond to physical properties, e.g., spin-up or spin-down, $|\uparrow\rangle, |\downarrow\rangle$, which depend on the choice of arbitrary measurement axes. In the context of gate-based quantum computing the orthogonal basis is labeled $|0\rangle, |1\rangle$ corresponding to classical computing bits. The label may also be used to indicate the data represented by a quantum state, which we discuss in Section 2.3 after introducing registers of qubits in Section 2.2.

## 2.2.    Registers of Qubits

In this section we consider a collection of qubits as a quantum register. In Section 2.4 we introduce operations on quantum registers.

**Definition 2**: A register of $n$ qubits is represented by a state vector $|\psi\rangle = |\psi_0\rangle \otimes |\psi_1\rangle \otimes ... \otimes |\psi_{n-1}\rangle \in \mathcal{H}$ where $\mathcal{H} \cong \mathcal{H}_0 \otimes \mathcal{H}_2 \otimes ... \otimes \mathcal{H}_n \cong \mathbb{C}^{2^{n-1}}$ and is described as a superposition of $2^n$ orthogonal states $|0\rangle, |1\rangle, ..., |2^n - 1\rangle$ as $|\psi\rangle = \sum\limits_{i=0}^{2^n-1} \alpha_i |i\rangle$ where $\sum\limits_{i=0}^{2^n-1} |\alpha_i|^2 = 1$.

A quantum register is a tuple of $n$ qubits [1], whose combined state is represented as the quantum system $|\psi\rangle \in \mathbb{C}^{2^n}$, corresponding to a tensor product of the individual Hilbert spaces of the qubits. The combined state $|\psi\rangle$ of two individual qubits $|\psi_1\rangle = \alpha_1|0\rangle + \beta_1|1\rangle$ and $|\psi_2\rangle = \alpha_2|0\rangle + \beta_2|1\rangle$ may be represented in Dirac notation as a Kronecker product, denoted by $\otimes$, of the individual qubits with various equivalent notations:

$$|\psi\rangle = |\psi_1\rangle \otimes |\psi_2\rangle = |\psi_1\rangle|\psi_2\rangle = |\psi_1\psi_2\rangle$$
$$= (\alpha_1|0\rangle + \beta_1|1\rangle)(\alpha_2|0\rangle + \beta_2|1\rangle)$$
$$= \alpha_1\alpha_2|0\rangle|0\rangle + \alpha_1\beta_2|0\rangle|1\rangle + \alpha_2\beta_1|1\rangle|0\rangle + \beta_1\beta_2|1\rangle|1\rangle$$
$$= \alpha_1\alpha_2|0\rangle \otimes |0\rangle + \alpha_1\beta_2|0\rangle \otimes |1\rangle + \alpha_2\beta_1|1\rangle \otimes |0\rangle + \beta_1\beta_2|1\rangle \otimes |1\rangle$$
$$= \alpha_1\alpha_2|00\rangle + \alpha_1\beta_2|01\rangle + \alpha_2\beta_1|10\rangle + \beta_1\beta_2|11\rangle$$
$$= \alpha_1\alpha_2 \begin{pmatrix} 1 \\ 0 \end{pmatrix} \otimes \begin{pmatrix} 1 \\ 0 \end{pmatrix} + \alpha_2\beta_1 \begin{pmatrix} 1 \\ 0 \end{pmatrix} \otimes \begin{pmatrix} 0 \\ 1 \end{pmatrix} + \alpha_1\beta_2 \begin{pmatrix} 0 \\ 1 \end{pmatrix} \otimes \begin{pmatrix} 1 \\ 0 \end{pmatrix} + \beta_1\beta_2 \begin{pmatrix} 0 \\ 1 \end{pmatrix} \otimes \begin{pmatrix} 0 \\ 1 \end{pmatrix} \qquad (3)$$
$$= \alpha_1\alpha_2 \begin{pmatrix} 1 \\ 0 \\ 0 \\ 0 \end{pmatrix} + \alpha_2\beta_1 \begin{pmatrix} 0 \\ 1 \\ 0 \\ 0 \end{pmatrix} + \alpha_1\beta_2 \begin{pmatrix} 0 \\ 0 \\ 1 \\ 0 \end{pmatrix} + \beta_1\beta_2 \begin{pmatrix} 0 \\ 0 \\ 0 \\ 1 \end{pmatrix} = \begin{pmatrix} \alpha_1\alpha_2 \\ \alpha_2\beta_1 \\ \alpha_1\beta_2 \\ \beta_1\beta_2 \end{pmatrix}.$$

Additional qubits will follow the same pattern, resulting in an exponentially large state space for the qubit register. A change in the order of the qubits in the quantum register simply shuffles the representation of the state corresponding to the definition of the Kronecker product.

Registers of qubits can be used to represent data in various formats, which is discussed in Section 2.3.

## 2.3.    Data Representation

Classical information is represented as bit strings in classical computers. This encoding scheme can also be used for quantum bits and is known as basis embedding. As an example, the bitstring '0101010010' may be represented as a quantum state $|0101010010\rangle$. However, this representation is inefficient since it requires the same number of qubits as classical bits and does not make use of the additional available degrees of freedom in quantum states: the phase and probability amplitudes of the individual basis states. As an example, a vector $\boldsymbol{a} \in \mathbb{R}^{2^n}$ will require $O(2^n)$ qubits to be approximately represented using amplitude encoding.

Embedding data into the phase of a basis state is known as phase or angle embedding. Similarly, embedding data into the probability amplitude of a basis state is known as amplitude encoding or amplitude embedding [17]. Both amplitude embedding and phase embedding are limited by the definition of quantum states, i.e., the normalization due to the Born rule and the periodicity of phases. Amplitude encoding is the method of choice for quantum algorithms for scientific computing and engineering [18].

As an example, consider the vector

$$\boldsymbol{a} = \Sigma_{i=0}^{2^{n}-1} a_i \boldsymbol{e}_i \qquad (4)$$

where $\boldsymbol{e}_i$ denotes the $i^{th}$ standard basis vector and $n$ is the number of qubits. If a vector has less than $2^n$ components, the remainder of the vector may be zero padded without any loss of generality. An amplitude encoding of this vector will be

$$|\boldsymbol{a}\rangle = \frac{1}{\|\boldsymbol{a}\|_2} \sum_{i=0}^{2^n-1} a_i |i\rangle \qquad (5)$$

where $|i\rangle$ is the $i^{th}$ basis vector of the quantum state. Note that a vector $\boldsymbol{a}$ in the ket denotes a quantum state encoding the entries of the vector in its probability amplitudes, while a scalar $i$ in the ket denotes the $i^{th}$ basis state of a quantum register, which is the standard basis vector $\boldsymbol{e}_i$.

In comparison, classical registers of bits occupy a combinatorial space with tuples of binary elements 0,1. A vector $\boldsymbol{a} \in \mathbb{R}^{2^n}$ is approximately represented on classical computers with $O(2^n)$ bits, e.g. using the IEEE-754 floating-point convention [19], while a quantum computer requires $n$ qubits using amplitude encoding, albeit as a normalized vector $\frac{\boldsymbol{a}}{\|\boldsymbol{a}\|}$.

In Dirac notation, an amplitude encoding of a vector $\boldsymbol{a}$ is typically denoted as $|\boldsymbol{a}\rangle$. The $i^{th}$ standard basis vector $\boldsymbol{e}_i$ in the computational basis is typically represented as $|i\rangle$.

### 2.4.    Gates

In this section, we introduce operations on the states of qubits or registers of qubits. In Section 2.5, we introduce operations to measure or read the states of qubits.

**Definition 3**: A quantum gate operating on $n$ qubits is represented as a unitary matrix $U \in \mathbb{C}^{2^n \times 2^n}$. In the gate-based quantum computing model operations on qubits are represented as quantum gates [1], or simply gates. Some quantum gates are analogous to classical gate operations, e.g. an $X$ gate is analogous to a classical $NOT$ operation. However, quantum gates may not have a corresponding classical counterpart, e.g. a Hadamard $H$ gate.

Quantum gates can be conveniently represented as complex unitary matrices. The simplest gates are single qubit gates represented as $SU(2) \in \mathbb{C}^{2 \times 2}$ matrices. Some important single-qubit gates are the identity gate $I$, the Pauli gate set $\{X, Y, Z\}$, and the Hadamard gate $H$. Their matrix representations are provided in Table 1:

*Table 1 Quantum gates with their matrix representations and descriptions*

| Gate | Matrix | Description |
| --- | --- | --- |

|  | Representation |  |
|---|---|---|
| $I$ | $\begin{pmatrix} 1 & 0 \\ 0 & 1 \end{pmatrix}$ | No operation on qubit. Qubit state is unchanged. |
| $X$ | $\begin{pmatrix} 0 & 1 \\ 1 & 0 \end{pmatrix}$ | Pauli X gate and matrix. $180°$ rotation around $X$. Transform basis states $|0\rangle$ to $|1\rangle$ and $|0\rangle$ to $|1\rangle$. |
| $Y$ | $\begin{pmatrix} 0 & -i \\ i & 0 \end{pmatrix}$ | Pauli Y gate and matrix. $180°$ rotation around $Y$. |
| $Z$ | $\begin{pmatrix} 1 & 0 \\ 0 & -1 \end{pmatrix}$ | Pauli Z gate and matrix. $180°$ rotation around $Z$. Apply phase of $-1$ to $|1\rangle$ |
| $H$ | $\frac{1}{\sqrt{2}}\begin{pmatrix} 1 & 1 \\ 1 & -1 \end{pmatrix}$ | $90°$ CW rotation around $X$. Used to create superposition from basis state $|0\rangle$. |

The $X$ gate is the quantum analog of the classical NOT gate. As an example, applying the $X$ gate to the basis state $|0\rangle$ yields $|1\rangle$:

$$X|0\rangle = \begin{pmatrix} 0 & 1 \\ 1 & 0 \end{pmatrix}\begin{pmatrix} 1 \\ 0 \end{pmatrix} = \begin{pmatrix} 0 \\ 1 \end{pmatrix} = |1\rangle. \tag{6}$$

Quantum gates can also act on $n$ qubits, in which case they can be represented in $SU(2^n) \in \mathbb{C}^{2^n \times 2^n}$. Multiple qubit gates typically arise as controlled versions of gates, of which $CNOT$ (or $cX$) is a commonly used one. A controlled gate manipulates the state of a "target qubit," conditioned on the state of a "control qubit." The $SWAP$ gate is another common gate that swaps the states between qubits. Controlled gates can be represented in block form. As an example, the matrix representation of the $cX$ gate is:

$$cX = |0\rangle\langle 0| \otimes I + |1\rangle\langle 1| \otimes X = \begin{pmatrix} I & \\ & X \end{pmatrix} = \begin{pmatrix} 1 & 0 & & \\ 0 & 1 & & \\ & & 0 & 1 \\ & & 1 & 0 \end{pmatrix} \tag{7}$$

where $|\cdot\rangle\langle\cdot|$ is an outer product.

As an example, outer products of basis states $|0\rangle$ and $|1\rangle$ operate on a qubit as:

$$(a|0\rangle\langle 0| + b|0\rangle\langle 1| + c|1\rangle\langle 0| + d|1\rangle\langle 1|)(\alpha|0\rangle + \beta|1\rangle) = \begin{pmatrix} a & b \\ c & d \end{pmatrix}\begin{pmatrix} \alpha \\ \beta \end{pmatrix}$$

$$= \begin{pmatrix} \alpha a + \beta b \\ \alpha c + \beta d \end{pmatrix} = (\alpha a + \beta b)|0\rangle + (\alpha c + \beta d)|1\rangle. \tag{8}$$

All quantum gates are unitary operations, which ensures normalization of quantum states according to the Born rule. Consequently, all quantum gates are also reversible operations, with the reverse operation simply being the Hermitian transpose or conjugate transpose of the quantum gate.

Gates acting on qubits corresponds to left multiplication, with the ket representing the quantum state of the qubits. As an example, consider a register of three qubits in the state $|000\rangle$. Applying a Hadamard gate to the first qubit (from the left) and a Pauli $X$ gate to the last qubit is represented as:

$$(H \otimes I \otimes X)|000\rangle = (H|0\rangle) \otimes (I|0\rangle) \otimes (X|0\rangle) = (\frac{1}{\sqrt{2}}|0\rangle + \frac{1}{\sqrt{2}}|1\rangle) \otimes (|0\rangle) \otimes (|1\rangle) =$$

$$\frac{1}{\sqrt{2}}(|001\rangle + |101\rangle) \tag{9}$$

As another example, consider a register of two qubits in the state $|00\rangle$ with a Hadamard gate applied to the first qubit followed by a $CNOT$ gate controlled by the first qubit applied to the second qubit:

$$(|00\rangle\langle00| + |01\rangle\langle01| + |10\rangle\langle11| + |11\rangle\langle10|)(H \otimes I)|00\rangle = |00\rangle + |11\rangle \tag{10}$$

Note that the order of operations progresses from right to left, which is the opposite of the order used for quantum circuits introduced in Section 2.6. It is customary not to include the $I$ gate when the qubits on which the operation is applied is implied.

Similar to classical Boolean logic, quantum gates can also form a universal gate set. This implies that any arbitrary quantum gate can be approximated using a finite sequence of gates from universal gate set. The Solovay-Kitaev theorem [20] is a central theorem in quantum computing which shows that the approximation error using a universal gate set scales as $O\left(log^c\left(\frac{1}{\epsilon}\right)\right)$ for a single-qubit gate where $c \approx 2$ and $O\left(mlog^c\left(\frac{m}{\epsilon}\right)\right)$ for a set of $m$ $CNOT$s and single-qubit unitaries. This corresponds to a polylogarithmic increase in the approximation using a universal gate set over the original number of arbitrary gates, which is efficient. The statement of the theorem is as follows:

**Theorem 4**: [21] Let $\mathcal{G}$ be an instruction set for $SU(2)$, and let a desired accuracy $\epsilon > 0$ be given. There is a constant $c$ such that for any $U \in SU(2)$ there exists a finite sequence $S$ of gates from $\mathcal{G}$ of length $O(\log^c(1/\epsilon))$ and such that $\|U - S\| \leq \epsilon$.

### 2.5. Measurement

In this section, we introduce the measurement operation to read out the states of qubits.

A quantum state is defined using probability amplitudes. To read a state, a series of measurements of the state needs to be performed, and the measurement statistics will correspond to the probability amplitudes of the quantum state and the basis used for the measurement [1].

We first define a measurement operator $M_m$ for measuring a quantum state $|\psi\rangle$.

**Definition 5**: Given a quantum state $|\psi\rangle$ and a measurement operator $M_m$ where $m$ is a measurement outcome, the probability of measuring outcome $m$ is

$$p(m) = \langle\psi|M_m^\dagger M_m|\psi\rangle \tag{11}$$

and the state of the quantum system after the measurement is

$$\frac{M_m|\psi\rangle}{\sqrt{\langle\psi|M_m^\dagger M_m|\psi\rangle}}. \tag{12}$$

such that the completeness relation

$$\sum_m p(m) = 1 \tag{13}$$

is satisfied.

In gate-based quantum computing projective measurements in the computational, or $|0\rangle$-$|1\rangle$, basis are typically used to sample quantum states which correspond to

$$M_0 = |0\rangle\langle0| = \begin{pmatrix} 1 & 0 \\ 0 & 0 \end{pmatrix}, M_1 = |1\rangle\langle1| = \begin{pmatrix} 0 & 0 \\ 0 & 1 \end{pmatrix} \tag{14}$$

and are represented by a "meter" symbol in quantum circuits. As an example, consider a two-qubit system in the state $\frac{1}{2}(|00\rangle + |01\rangle + |10\rangle + |11\rangle)$, with a measurement being performed on the first qubit. The first qubit is measured in the state $|0\rangle$ with $p(0) = \frac{1}{2}$ and the final state is:

$$(|0\rangle\langle0| \otimes I)(\tfrac{1}{2}(|00\rangle + |01\rangle + |10\rangle + |11\rangle)) = \tfrac{1}{\sqrt{2}}(|00\rangle + |01\rangle). \tag{15}$$

Similarly, the first qubit is measured in the state $|1\rangle$ with $p(1) = \frac{1}{2}$ and the final state is:

$$(|1\rangle\langle1| \otimes I)(\tfrac{1}{2}(|00\rangle + |01\rangle + |10\rangle + |11\rangle)) = \tfrac{1}{\sqrt{2}}(|10\rangle + |11\rangle). \tag{16}$$

Measurements may be performed to either read out an entire quantum register or as a flag to indicate successful operation by measuring ancilla qubits (introduced in Section 2.9).

Note that measurements will provide the squared moduli (probability) of the complex numbers defining the probability amplitude of a quantum state. In order to recover additional information, like the arguments (phases), a process known as quantum state tomography [22] is performed. Quantum states also have an overall phase that is not measurable; only the relative overall phase between two states is measurable.

Measurements are typically performed in the $Z$ or $|0\rangle$-$|1\rangle$ basis. A hardware implementation will typically only allow a $Z$ basis measurement. To perform measurements in another basis, the qubits can be 'rotated' to the desired basis by applying gate operations and then measured.

## 2.6.    Circuits

In this section, we introduce the circuit representation of quantum computing. We then demonstrate the properties of superposition and entanglement with their corresponding circuits in Sections 2.7 and 2.8.

Working with algebraic forms of algorithms can be unwieldy and difficult to visualize. Quantum circuits are a convenient representation of qubits and the operation sequence.

As an example, consider a register of three qubits $|abc\rangle$, with an $H$ gate applied to the qubit $a$ and an $X$ gate applied to the qubit $c$, followed by a $CNOT$ gate controlled by $a$ applied to $b$, and finally, measurements on all the qubits. These operations can conveniently be represented in circuit form as shown in Figure 1.

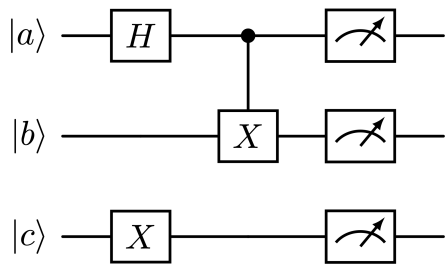

*Figure 1: Quantum circuit with gates and measurements.*

Note that the operations are ordered from left to right, unlike operations ordered right to left in algebraic notation. A controlled operation can be applied conditioned on the control qubit being either in the state $|1\rangle$ or $|0\rangle$. This in indicated in a quantum circuit with a filled or empty circle respectively, as shown in Figure 2.

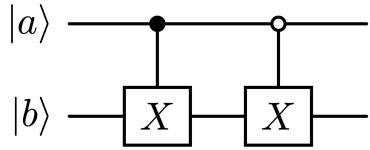

*Figure 2:* ***CNOT*** *gates conditioned on qubit **A** being in states **1** and **0**.*

Quantum registers can be ordered using either little-endian or big-endian notation. Little-endian ordering corresponds to the right-most qubit in algebraic notation indexed as 0 and big-endian corresponds to the left-most qubit indexed as 0. In a quantum circuit the first qubit from the top in a register is indexed as 0. When implementing quantum algorithms, the endian notation must be kept consistent since a change in ordering results in a shuffling of the basis states. The shuffling leads to the measurement distribution being distorted, leading to an incorrect interpretation of the measurement distribution. Both big- and little-endian conventions are used in libraries for quantum computing. As an example, Qiskit [23] uses little-endian notation, while Pennylane [24] uses big-endian notation. We repeat the example in Figure 1 but with $|abc\rangle$ represented in little-endian notation. In this paper, we use big-endian notation for all figures and equations and Qiksit code implementations are in little-endian notation. We provide an example code with the corresponding circuit in Figure 3:

```
#!/usr/bin/python3

from matplotlib import pyplot as plt
import qiskit

# Create a register of 3 qubits
myQRegister = qiskit.QuantumRegister(3, '\psi')

# Create a register of 3 classical bits
myCRegister = qiskit.ClassicalRegister(3,'Classical Bits')

# Create a quantum circuit using myRegister
myCircuit = qiskit.QuantumCircuit(myQRegister, myCRegister)

# Start working in big-endian order

# Add the gates in order of operation
myCircuit.h(0)
myCircuit.x(2)
myCircuit.cx(0,1,ctrl_state='1')
```

```
# Insert a barrier to keep circuit organized
myCircuit.barrier()

# Measure all the qubits in myQRegister and store state in myCRegister
myCircuit.measure(myQRegister,myCRegister)

# Convert from big-endian to little-endian by reversing bits
myCircuit = myCircuit.reverse_bits()

# Draw the circuit
myCircuit.draw('mpl')
plt.show()
```

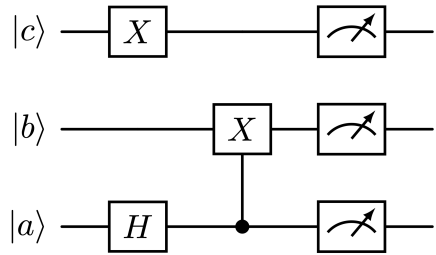

*Figure 3 The circuit in Figure 1 in little endian convention.*

## 2.7.    *Superposition*

Superposition is a unique property of quantum systems [1]. As an example, a single qubit $|\psi\rangle$ can exist in a superposition of $|0\rangle$ and $|1\rangle$. A register of $n$ qubits can exist in an exponentially large superposition of $N = 2^n$ states, which form an orthonormal basis of the Hilbert space $\mathcal{H}: \mathbb{C}^{2^n}$. Representation of an arbitrary quantum state of $n$ qubits requires $O(2^n)$ classical registers. This compact representation as a superposition of basis states enables exponentially efficient representation of data on quantum computers.

The 0-1 basis vectors are assigned to the $Z$ basis vectors of qubits by convention. Hardware implementations of qubits will typically allow measurement in only one basis, e.g. the spin state of a fermion. To measure in another basis in hardware implementations it is convenient to rotate the qubits and measure them instead of implementing measurement in a rotated basis. However, single qubit rotations do not correspond to arbitrary unitary rotations. Instead, they are limited by the Kronecker product structure of the qubits. An arbitrary rotation in $\mathcal{H}: \mathbb{C}^{2^n}$ involves multiple-qubit gates and entanglement, which is discussed in Section 2.8.

The Hadamard ($H$) gate puts qubits in uniform superposition, in which all basis states have the same probability amplitude. As an example, we provide code below to put qubits in uniform superposition and measure their states as shown in the quantum circuit in Figure 4 with the output shown in Figure 5.

The Hadamard gate is applied to all three qubits to put them in the uniform superposition state

$$(H \otimes H \otimes H)|000\rangle = \frac{1}{2\sqrt{2}}(|000\rangle + |001\rangle + |010\rangle \ldots |111\rangle). \tag{17}$$

All three qubits are then measured, which yields one of the basis states $|000\rangle, |001\rangle, |010\rangle, \ldots, |111\rangle$. This process is repeated 32768 times to obtain the distribution shown in Figure 5. As the number of measurements is increased, the distribution of measurements for qubits in uniform superposition converges to the uniform distribution.

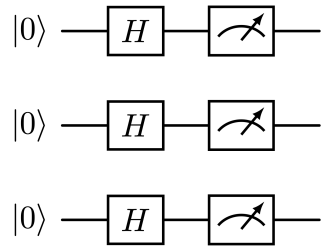

*Figure 4 Circuit for uniform superposition.*

```
#!/usr/bin/python3
from matplotlib import pyplot as plt
import qiskit
from qiskit_aer import Aer
from qiskit.visualization import plot_histogram

# Create a register of 3 qubits
myQRegister = qiskit.QuantumRegister(3, '\psi')

# Create a register of 3 classical bits
myCRegister = qiskit.ClassicalRegister(3,'Classical Bits')

# Create a quantum circuit with using myRegister
myCircuit = qiskit.QuantumCircuit(myQRegister, myCRegister)

# Hadamard gates on al qubits
myCircuit.h(myQRegister)

# Measure all the qubits in myQRegister and store state in myCRegister
myCircuit.measure(myQRegister,myCRegister)

# Simulate the circuit
mySimulator = Aer.get_backend('aer_simulator')
```

```
result = mySimulator.run(myCircuit,shots=2**15).result()

# Plot a bar chart of all the results
plot_histogram(result.get_counts(), title='Uniform Superposition')

plt.show()
```

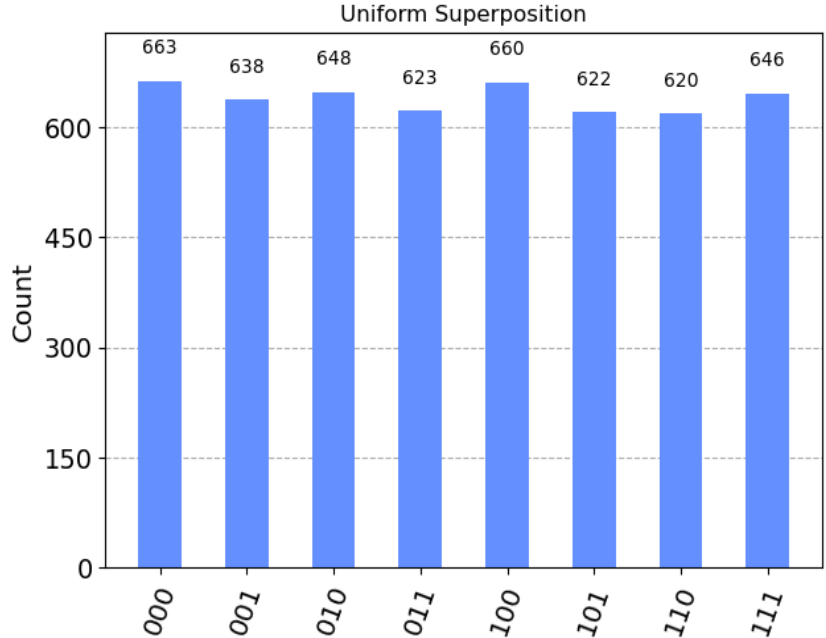

*Figure 5 Distribution of measurements for qubits in uniform superposition*

## **2.8.    Entanglement**

Entanglement is another property unique to quantum systems [1]. Entangled qubits have a correlated state. A maximal entanglement for a pair of qubits means that the state of one qubit can be fully determined by measuring the state of the other qubit. The simplest set of maximally entangled states are the Bell states:

$$|\Phi^+\rangle = \tfrac{1}{\sqrt{2}}(|00\rangle + |11\rangle), |\Phi^-\rangle = \tfrac{1}{\sqrt{2}}(|00\rangle - |11\rangle),$$
$$|\Psi^+\rangle = \tfrac{1}{\sqrt{2}}(|01\rangle + |10\rangle), |\Psi^-\rangle = \tfrac{1}{\sqrt{2}}(|01\rangle - |10\rangle). \tag{18}$$

As an example, if the first qubit of the state $|\Phi^+\rangle = \tfrac{1}{\sqrt{2}}(|00\rangle + |11\rangle)$ is measured as 0, the second qubit is also in the state 0, and if the first qubit is measured as 1, the second qubit is also in the state 1. Note that this is different from the state $|\psi\rangle = |00\rangle$, which is not an entangled state.

Mathematically, entangled qubits cannot be expressed as a Kronecker product of individual qubit states. The state $|00\rangle$ can be written as $|0\rangle \otimes |0\rangle$, however, it is impossible to separate $|\Psi^+\rangle$ into

such a Kronecker product of Z basis states, or any other basis formed by a tensor product of the bases individual qubits. Similarly, single-qubit gates cannot rotate an entangled quantum state to a state separable into Kronecker products.

Entangled states are necessary for quantum registers to go from a Hilbert space described by individual $SU(2)$ rotations of qubit states to the larger Hilbert space described by $SU(2^n)$ rotations of a quantum register.

In Figure 6, we provide an example of a circuit creating the entangled state $|\Phi^+\rangle$ with the output measurements shown in Figure 7. Code to create and simulate this circuit is provided below.

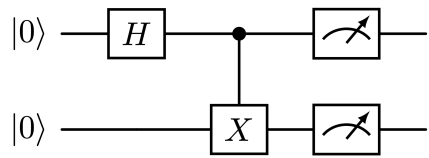

*Figure 6 Circuit for Bell state $|\Phi^+\rangle$*

```
#!/usr/bin/python3

from matplotlib import pyplot as plt
import qiskit
from qiskit_aer import Aer
from qiskit.visualization import plot_histogram

# Create a register of 2 qubits
myQRegister = qiskit.QuantumRegister(2, '\psi')

# Create a register of 2 classical bits
myCRegister = qiskit.ClassicalRegister(2,'Classical Bits')

# Create a quantum circuit with using myRegister
myCircuit = qiskit.QuantumCircuit(myQRegister, myCRegister)

# Hadamard gates on first qubit
myCircuit.h(0)
# CNOT gate controlled by first qubit on second qubit
myCircuit.cx(0,1)

# Measure all the qubits in myQRegister and store state in myCRegister
myCircuit.measure(myQRegister,myCRegister)

# Simulate the circuit
mySimulator = Aer.get_backend('aer_simulator')
result = mySimulator.run(myCircuit,shots=5120).result()

# Plot a bar chart of all the results
plot_histogram(result.get_counts(), title='Bell State')
```

```
plt.show()
```

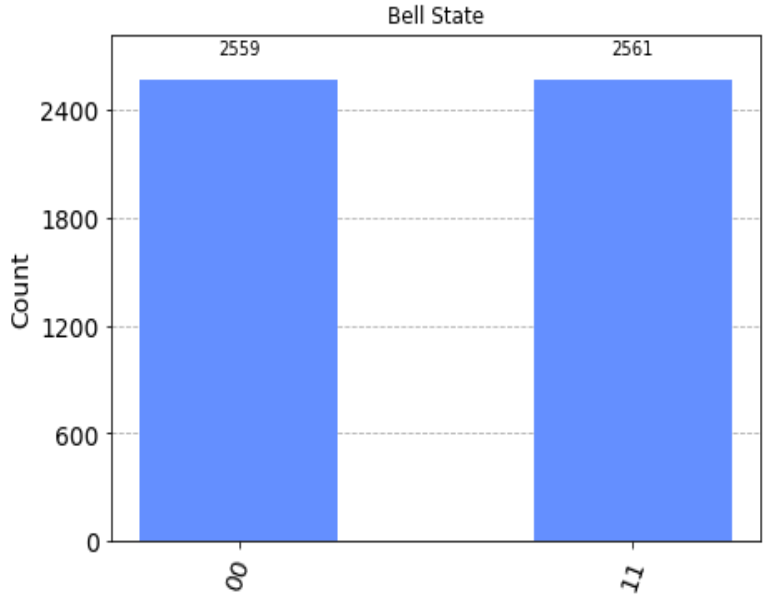

*Figure 7 Measurement distribution for entangled qubits*

## 2.9.    *Classical and Reversible Computation*

All quantum gates are reversible since they are unitary operations. Therefore, the corresponding inverse operation of a gate $O$ is $O^{\dagger}$, where † denotes the Hermitian conjugate of $O$. For a quantum circuit, without measurement, the inverse operation is application of the Hermitian conjugates of the gates in reverse order.

However, classical computation operations are not necessarily reversible, e.g. the classical logic gates AND and OR. Regardless, it is possible to perform all classical computation operations on quantum computers [1].

First, we note that the classical NAND gate is a universal classical gate, i.e. any classical Boolean operation can be represented as a combination of NAND gates. We also note that the FANOUT operation may be needed in classical logic circuits involving combinations of NAND gates. Therefore, if the NAND operation and FANOUT operation can be performed on a gate-based quantum computer then any classical computation can be performed on a quantum computer.

A reversible version of these gates can be formed by using ancillary qubits, or "ancilla" qubits. Ancilla qubits are often utilized in quantum computing to implement nonunitary or irreversible operations. The measured state of ancilla qubits can either be discarded at the end of the computation, used to "postselect" the unmeasured state in the remaining qubits (e.g. continue if ancilla is measured as $|1\rangle$, restart if measured as $|0\rangle$), or used for branching operations for the remainder of a quantum-classical workflow.

Even though this approach does not make efficient use of quantum resources, it implies that all classical computation can be performed on quantum computers. Both the NAND and FANOUT operations can be performed using a Toffoli gate, or a *ccNOT* gate as shown in Figure 8. Note that measurements are not reversible since they are projection operations.

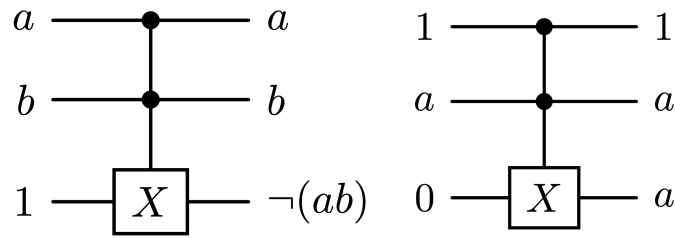

*Figure 8 A quantum implementation of a NAND gate and a FANOUT operation*

### 2.10.    Access Models

Access models in quantum computing define how data are accessed and manipulated within quantum algorithms. Efficient data access is crucial for the performance and scalability of these algorithms. Two primary access models are commonly used for matrix operations: the sparse access model and the block-encoding model (also known as qubitization). These models provide different approaches to handling matrices, optimizing for either sparsity or unitary representation, respectively.

In these access models, quantum oracles play a fundamental role. Oracles are unitary operations that act as "black boxes," encapsulating specific operations or data needed by the algorithm. In the case of a linear system problem $Ax = b$ this could be the entries of the matrix $A$ or a method to prepare the vector $b$. The complexity of quantum algorithms is often defined by the number of oracle accesses required, as these oracles abstract the implementation details of complex operations, allowing algorithms to access necessary information efficiently. The number of accesses or "queries" to oracles is often referred to as the query complexity of a quantum algorithm. A typical underlying assumption is that the information being accessed using the oracle is efficiently computable. For most algorithms, an oracle implementation that scales polylogarithmically in the number of gates with the problem size (or polynomially with the number of qubits) is considered efficient.

There are a variety of different oracles for various operations in quantum algorithms. An early access model for matrices, on which the HHL quantum linear system algorithm is based, is access to an approximation of the unitary matrix operator $U \approx e^{itA}$, where $A$ is a Hermitian matrix, through a. Hamiltonian simulation procedure (Section 5.1). For the problems considered in this paper the more recent and efficient sparse access model [5] and block-encoding model [25], [26] are used to access the entries of a matrix, and a state preparation oracle is used to prepare a quantum state used by an algorithm.

The implementation of efficient oracles for practically relevant problems is an active area of research [27]. An example of a circuit using a sparse matrix access model to encode $U \approx e^{iA}$ where $A = \sum_j A_j$ is a tridiagonal Toeplitz matrix decomposed as a sum of 1-sparse matrices $A_j$ can be found in [28]. [29] and [30] provide access models for discrete Laplacian matrices.

We introduce the Sparse Access Model and the Block-encoding model for matrices in Sections 2.10.1 and 2.10.2, respectively. In Section 2.10.3 we introduce the Hermitian dilation trick to encode an arbitrary matrix as a Hermitian matrix. We then provide a recipe to block-encode oracles Hermitian matrices in Section 2.10.4.

### 2.10.1.    Sparse Access Model

The sparse access model is designed for matrices with a high proportion of zero entries. It uses position and value oracles (unitary operations) to efficiently retrieve non-zero elements:

- **Position Oracle $\mathcal{O}_A^{pos}$**: Provides the position of non-trivially zero elements in a row.

    $$\mathcal{O}_A^{pos}: |i, v\rangle \rightarrow |i, j(i, v)\rangle \ \forall \ i, j \in \{1, \dots, N\}, v \in \{1, \dots, d\}$$

$i$: row (or column) number

$j$: column (or row) number

$v$: enumeration of (typically) non-zero entries in row (or column) $i$

- **Value Oracle $\mathcal{O}_A^{val}$**: Provides the value of matrix elements.

$$\mathcal{O}_A^{val}: |i, j, z\rangle \rightarrow |i, j, A_{i,j} \oplus z\rangle \ \forall \ z \in \{0,1\}^{\otimes s}$$

$z$: initialized bitstring of length $s$

$A_{i,j} \oplus z \in \{0,1\}^{\otimes s}$: a bitstring of length $s$ encoding the matrix entry $A_{i,j} \in \mathbb{C}$.

A more powerful and general access model is the block-encoding model.

## 2.10.2.    Block-Encoding Model

Another access model is the block-encoding oracle which is access to a unitary (any quantum circuit) of the form

$$U_A = \begin{pmatrix} A/\alpha & * \\ * & * \end{pmatrix} \tag{19}$$

where $\alpha$ is a subnormalization factor and $*$ are irrelevant entries.

This form is central to methods based on qubitization and quantum signal processing, which are the most powerful and optimal methods for most problems. The main idea is to apply the block-encoded unitary to a state

$$U_A|0\rangle|\boldsymbol{b}\rangle = \begin{pmatrix} \frac{A}{\alpha} & * \\ * & * \end{pmatrix} \begin{pmatrix} \boldsymbol{b} \\ 0 \end{pmatrix} = \begin{pmatrix} \frac{A\boldsymbol{b}}{\alpha} \\ * \end{pmatrix} = \frac{1}{\alpha}|0\rangle|A\boldsymbol{b}\rangle + |\perp\rangle \tag{20}$$

Measuring the first qubit in the state $|0\rangle$ indicates a successful matrix-vector multiplication. Note that although this $2 \times 2$ block encoding uses a single ancilla qubit, a register of qubits can also be used for block encoding which requires $m$ ancilla qubits to be measured as $|0\rangle^{\otimes m}$. We also note that since $U_A$ is unitary, it is necessary that $\|A/\alpha\|_2 \leq 1$. The success probability of measuring the ancilla qubits as $|0\rangle^{\otimes m}$ can be defined as follows [18]: Consider any $\alpha$ such that $\|A/\alpha\|_2 \leq 1$ so that

$$A/\alpha = (\langle 0^m| \otimes I_n)U_A(|0^m\rangle \otimes I_n) \tag{21}$$

The probability of successfully measuring $|0\rangle^{\otimes m}$ is

$$p(|0\rangle^{\otimes m}) = \frac{1}{\alpha^2} \| A|\boldsymbol{b}\rangle \|^2 \tag{22}$$

Block-encoded oracles also allow addition, products, and tensor products of matrices. [25], [31], [32] provide implementation details of these operations. The specifics of oracle implementations for access models are beyond the scope this review since they are problem-dependent and an active area of research. Recent work has developed circuits to encode various classes of sparse matrices [27], [33].

A sparse access model can be transformed into a block-encoding model using $O(1)$ queries to $\mathcal{O}_A^{pos}$ and $\mathcal{O}_A^{val}$ and $O(\text{poly} \log n)$ additional gates [34].

### 2.10.3. Hermitian Dilation

We note that various formalisms in quantum computing including oracles for matrices may require the matrices to be Hermitian. In case a matrix $\boldsymbol{A}$ is not Hermitian, its Hermitian dilation $\boldsymbol{H}$ can often be used instead:

$$H = \begin{pmatrix} 0 & A \\ A^\dagger & 0 \end{pmatrix}. \tag{23}$$

The Hermitian dilation has the same condition number as $\boldsymbol{A}$, is diagonalizable, and the eigenvalues $\lambda_i$ of the $\boldsymbol{H}$ are pairwise $\pm\sigma_i$, the singular values $\sigma_i$ of the matrix $\boldsymbol{A}$. Therefore, any guarantee of positive- or negative-definiteness of the matrix $\boldsymbol{A}$ is lost. This is apparent by rewriting using the singular value decomposition (SVD) of $\boldsymbol{A}$:

$$\begin{pmatrix} 0 & A \\ A^\dagger & 0 \end{pmatrix} = \begin{pmatrix} 0 & U\Sigma V^\dagger \\ V\Sigma U^\dagger & 0 \end{pmatrix} = \tfrac{1}{2} \begin{pmatrix} U & -U \\ V & V \end{pmatrix} \begin{pmatrix} \Sigma & 0 \\ 0 & -\Sigma \end{pmatrix} \begin{pmatrix} U^\dagger & V^\dagger \\ -U^\dagger & V^\dagger \end{pmatrix}. \tag{24}$$

### 2.10.4. Pauli Basis and Decomposition

One may always implement an oracle for a matrix $\boldsymbol{A} \in \mathbb{C}^{2^n \times 2^n}$ by decomposing $\boldsymbol{A}$ in the Pauli basis. First, we note that any Hermitian matrix $\boldsymbol{A} \in \mathbb{C}^{2 \times 2}$ may be represented as a linear combination of Pauli matrices, denoted as $\sigma_x = X, \sigma_y = Y, \sigma_z = Z$ and $\sigma_I = I$ where $X, Y, Z$ and $I$ are single qubit gates shown in Table 1, as

$$A = \alpha_1 \sigma_I + \alpha_2 \sigma_x + \alpha_3 \sigma_y + \alpha_4 \sigma_z. \tag{25}$$

where $\alpha_i \in \mathbb{C}$. This can be extended to a general matrix $\boldsymbol{A} \in \mathbb{C}^{2^n}$ as the sum

$$\boldsymbol{A} = \frac{1}{2^n} \sum_{i_1, i_2, \ldots, i_n} \alpha_{i_1, i_2, \ldots, i_n} \, \sigma_{i_1} \otimes \sigma_{i_2} \otimes \ldots \otimes \sigma_{i_n} \tag{26}$$

where $\alpha_{i_1, i_2, \ldots, i_n} \in \mathbb{C}$ are the coefficients of $\boldsymbol{A}$ in the Pauli basis and $\sigma_{i_j} \in \{I, X, Y, Z\}$ are the identity and Pauli matrices. The coefficients $\alpha_{i_1, i_2, \ldots, i_n}$ can be computed as

$$\alpha_{i_1, i_2, \ldots, i_n} = Tr\left(\boldsymbol{A} \odot \left(\sigma_{i_1} \otimes \sigma_{i_2} \otimes \ldots \otimes \sigma_{i_n}\right)\right) \tag{27}$$

where $Tr(\cdot)$ is the trace of a matrix, and $\odot$ denotes the Hadamard product (element-wise multiplication) of the matrix entries.

A term of the form $\sigma_{i_1} \otimes \sigma_{i_2} \otimes \ldots \otimes \sigma_{i_n}$ is often referred to as a Pauli string and written as $\sigma_{i_1} \sigma_{i_2} \ldots \sigma_{i_n}$.

The linear combination of unitaries (LCU) subroutine, described in Section 4.1.5 can be used to implement a linear combination of Pauli basis terms [29]. In fact, LCU implementations produce a block-encoded oracle [18], [25]. Although $d$ Pauli basis terms produce at most a $d$-sparse matrix, i.e., a matrix with at most $d$ non-zero entries in any row or column, a $d$-sparse matrix in general does not correspond to at most $d$ Pauli basis terms [27]. We provide as an example a code in Appendix A.1 to decompose the Laplacian matrix into its Pauli basis and show the growth of the number of terms with $N$ in Figure 9.

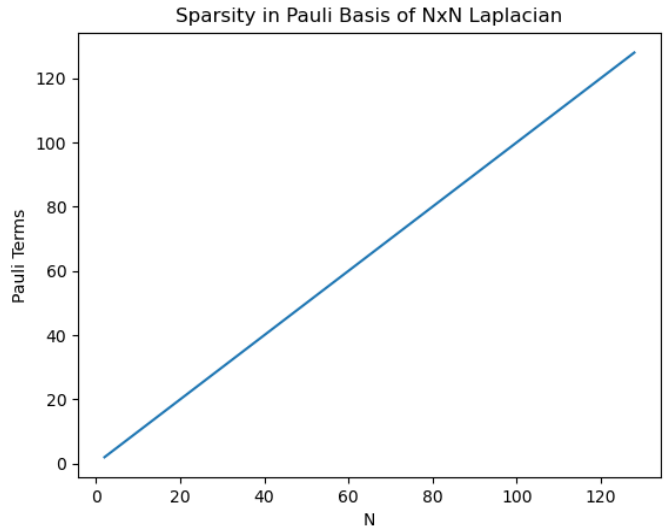

*Figure 9 Growth in the number of Pauli basis terms for a 1D Laplacian matrix*

## 2.11.  Limitations

Quantum physics places some fundamental limits on possible operations using quantum computing. A well-known limitation is the no-cloning theorem [35], which states that an arbitrary quantum state cannot be used to make an exact, independent copy of itself. More precisely, the map $U|\phi\rangle|\psi\rangle \rightarrow e^{i\alpha}|\psi\rangle|\psi\rangle$ is not possible in general for arbitrary $|\psi\rangle \in \mathbb{C}^n$ and $|\phi\rangle \in \mathbb{C}^n$ where $U \in \mathbb{C}^{2n}$.

The no-deletion theorem complements the no-cloning theorem, which prohibits information in a quantum state from being erased (using unitary operations). Note that imperfect copies are still possible to create, with known bounds on the error [36], [37], [38], or perfect copies can be made if the quantum state is fully known.

The normalization of a state and the periodicity of phase can also be considered limitations on the state. All gate operations are unitary and linear. To apply nonunitary and nonlinear operations a projection onto a subspace of the overall linear space must be considered. This is the central idea behind block-encoding. This property is exploited in various quantum computing algorithms, at the expense of ancilla qubits and a non-zero probability of failure [39].

Amplifying the probability of a desirable quantum state amongst a superposition of undesirable states can also pose a challenging limitation. If the amplitude of the desired state is exponentially small, the probability of obtaining the desired state cannot be boosted without an exponential overhead [40]. This is known as the post-selection problem.

Getting data in and out of a quantum register is also a challenging problem. I/O is expensive on quantum computers: preparing or reading out an arbitrary quantum state scales as $O(2^n)$. Furthermore, reading out a quantum state with the phases requires quantum state tomography [41]. The quantum version of random-access memory, QRAM, has been proposed to efficiently access and store data on quantum computers and is an active area of research [42].

Some problems have been proven to exhibit no quantum speedup compared to classical computing. A well-known result is the quantum no-fast-forwarding theorem [43] which states that the optimal scaling for an arbitrary Hamiltonian simulation is $O(T)$. The proxy problem for the proof is computing the parity of a string of bits. Classical methods query the string $N$ times, whereas the optimal quantum algorithm will query an oracle for the string $N/2$ times [44]. However, for special cases, e.g. Hamiltonian simulation of positive-definite systems, fast-forwarding is possible [39].

Despite these limitations, there is potential for quantum computing to make an impact on scientific computation and engineering problems due to the parallelism enabled by quantum computers and the exponentially large state spaces of quantum registers. Classical computers and algorithms are simply not capable of storing and processing large quantum simulations, making quantum computers the only viable option for general quantum simulation problems [2] despite the no-fast-forwarding theorem. Algorithms typically need to be modified to be amenable to quantum computing. For example, [45] uses a Carleman linearization of a nonlinear ordinary differential equation to obtain an exponential speedup in the number of unknowns, and [46] presents a novel data encoding scheme for the efficient implementation of a quantum lattice Boltzmann method.

## 3. Programming Quantum Computers

In this section, we provide a high-level overview of the programming stack for quantum computing, with examples of various libraries and implementations. We first provide an overview of the quantum computer programming stack for gate-based quantum computers in Figure 10.

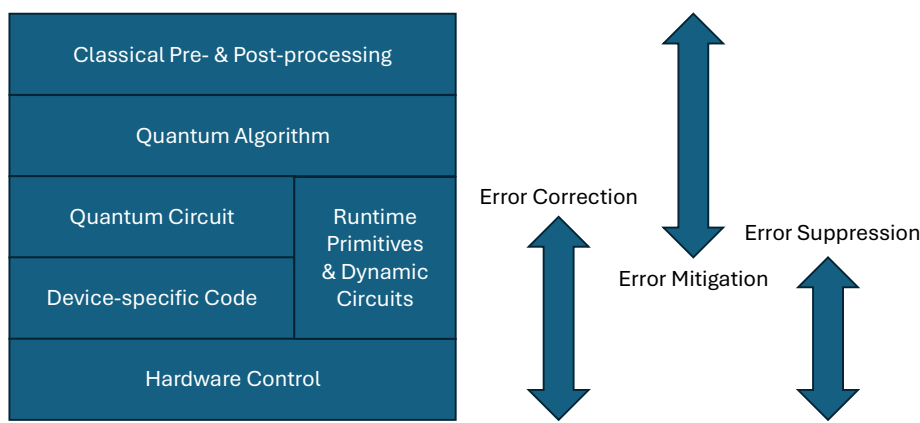

*Figure 10 A quantum computer programming stack for gate-based quantum computers.*

The hardware-level controls are specific to the particular implementation of the devices. For superconducting qubits, this is typically the control and shaping of microwave pulses to apply quantum gates and perform measurements [47]. For a photonic quantum computer, this can be an implementation of a phase shifter or a beam splitter [48]. This typically requires detailed knowledge of the physics of hardware implementation. Quantum circuits allow the abstraction of these details up to at least the hardware gate-set by manufacturers of devices. However, libraries for quantum circuits typically provide access to these low-level controls.

The device-specific code includes the gate set of the hardware and runtime primitives. Runtime primitives are quantum instructions conditioned on qubit measurements (classical) within a quantum circuit that are executed at circuit runtime and typically include quantum state sampling, expectation value estimation, and dynamic circuits. As an example, using dynamic circuits, if a qubit is measured as 0 during circuit execution, certain gates may be applied; otherwise, measuring 1 leads to other gates being applied.

A quantum assembly language (QASM) is a bridge between device-specific code and the hardware-level implementation of the circuit. OpenQASM [49], cQASM [50], and Jaqal [51] are well-known examples of QASMs. OpenQASM is an open-source language developed by researchers at IBM and the most widely used one in libraries for quantum computing.

A code describing a quantum circuit may be defined using an arbitrary set of gates, which can then be transpiled to code for a target device. In addition to converting to the gate set of the device, the topology of the device must also be considered since logical qubits in the quantum circuit need to be mapped to physical qubits (or a collection of physical qubits corresponding to a logical qubit) in the device. This process is typically executed by a high-level language or library and is referred to as a layout and routing for the tasks of mapping algorithm qubits to physical qubits and introducing SWAP operations to enable interactions between qubits that are not physically connected, respectively.

There are various high-level libraries developed for quantum computing, of which the vast majority are available in Python 3. Currently, the most advanced and stable library suite is the open-source library Qiskit [23], developed by IBM. Google has also initiated its efforts in implementing its library Cirq [52]. However, it is nascent and expected to undergo major changes before a stable version is made available. Microsoft, on the other hand, has developed Q# [53], a new language specific to quantum computers as a bottom-up approach. Other notable libraries are Pennylane [24] and Strawberry Fields [54], developed by Xanadu for their photonic quantum computers, and PyQuil [55] for Rigetti's superconducting qubit architecture. Recently, MATLAB has introduced a quantum computing toolbox for constructing simple quantum circuits and simulating or submitting them to systems available through the cloud. However, it is in a nascent stage and does not provide the high-level functionality available in the Qiskit, Cirq, or Pennylane libraries.

Several of the high-level libraries have built-in implementations of algorithmic primitives. There are also additional wrappers for the implementation of algorithms for machine learning and quantum chemistry. Tensorflow quantum is a wrapper for Cirq and Tensorflow for quantum machine learning and variational quantum algorithms. Qiskit also provides tools for hardware design in the Qiskit-metal library [23].

There are several important libraries for the classical preprocessing steps for quantum computation. OpenFermion [56], PySCF [57] and Qiskit-Nature can be used for quantum chemistry simulation preprocessing, e.g., for selecting basis sets for molecular structure or implementations of Jordan-Wigner [58] or Bravyi-Kitaev [59] transformations. Quantum signal processing algorithms require a sequence of phase angles to be computed classically. The QSPPACK [60] and PyQSP [26] libraries are the only libraries for this task as of now, with QSPPACK providing superior performance and implementations of state-of-the-art algorithms for phase factor calculations. Since classical preprocessing libraries can vary widely by application area, it is beyond the scope of this paper to provide a thorough review of these libraries.

Finally, we note that quantum error correction, mitigation, and suppression can be part of various levels of the quantum computing stack. For classical computers, error correction and fault-tolerance typically occur at the bit level, e.g., voting circuits for logic, redundancy for memory, and parity for communications. In contrast, present-day error correction for quantum computing ranges from preprocessing steps of the algorithm to hardware-level controls of individual qubits [61]. Control hardware typically needs to be calibrated to counteract noise and fabrication defects. The prevalent method for constructing a logical qubit is centered around error correction codes running on a multitude of physical qubits, which run at the device-specific code level. Error mitigation techniques like reordering or optimizing circuits can occur at the quantum circuit or device-specific code levels [62]. For NISQ algorithms, techniques like zero-noise extrapolation are used at the quantum algorithm and pre- and post-processing levels to reduce the effects of noise on measured observables [63].

Quantum computers are typically accessed through the cloud. Several commercial, academic, and government research groups have made their prototypes available either directly or through a third party. Notable commercial vendors are IBM, Google, IonQ, Honeywell, Xanadu, and Rigetti, with Amazon bra-ket and Microsoft Azure providing third-party cloud access. Quantum hardware testbeds are available at Lawrence Berkeley National Laboratory (AQT) and Sandia National Laboratories (QSCOUT). A compute job for sampling quantum states consists of the circuit and parameters, if any, along with a mapping of the qubits in the circuit to physical qubits in hardware. Since the Qiskit library suite encompasses the entire quantum computer programming stack for universal or gate-based quantum computing, is well-established, and is not expected to undergo major changes, it is our choice of library for programming quantum computers.

## *4. Important Subroutines*

Since the quantum computational model differs significantly from classical computing, familiar and routinely used mathematical operations in scientific computing need to be recast in a form amenable for quantum computers. Although a plethora of subroutines and operations have been developed, we limit our discussion to subroutines that appear in the algorithms in Section 5.

Figure 11 provides a high-level overview of Sections 4 and 5.

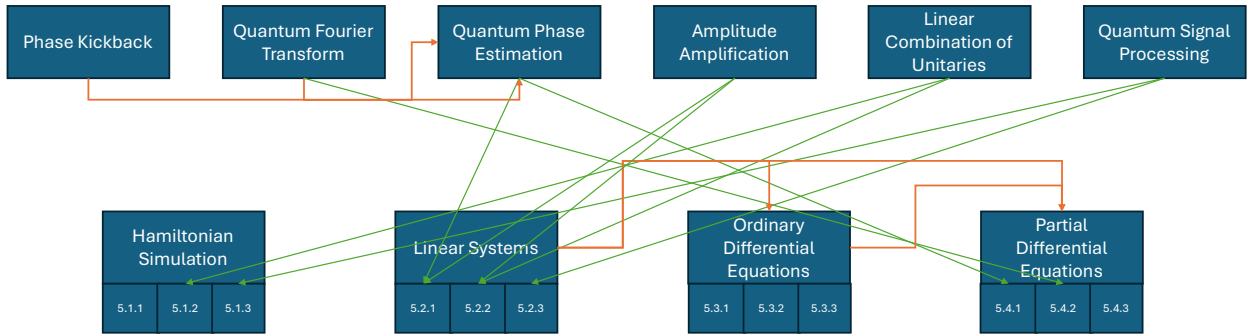

*Figure 11 Overview of subroutines and algorithms presented in this paper.*

### 4.1.1.     **Phase Kickback**

In classical logic gate computational model, the control bits always remain unchanged. This does not hold true for the quantum computational model. This phenomenon is exploited in phase kickback, which can be utilized for operations related to the eigendecomposition of operations on a quantum computer.

Phase kickback is a uniquely quantum phenomenon where the control qubit in a controlled operation is affected by the operation, whereas the target qubit remains unchanged. Whenever a controlled gate is applied, the eigenvalues of the gate can be 'kicked back' to the phase of the control qubit [1].

Consider a unitary operation in its eigenbasis $U = \sum_i e^{i\lambda_i} |\lambda_i\rangle \langle\lambda_i|$ with eigenvectors $|\lambda_i\rangle$ and eigenvalues $e^{i\lambda_i}$. The controlled version of this operation, controlled by one qubit, can be written as:

$$cU = |0\rangle\langle0| \otimes I + |1\rangle\langle1| \otimes \sum_i e^{i\lambda_i} |\lambda_i\rangle\langle\lambda_i| \tag{28}$$

where the first qubit is the control qubit.

Now consider a quantum state $|\psi_i\rangle = \frac{1}{\sqrt{2}} (|0\rangle + |1\rangle) \otimes |\lambda_i\rangle$, where the control qubit is in uniform superposition, and the target qubit register is in an eigenstate $|\lambda_i\rangle$ of $U$. Applying the controlled unitary operation $cU$ to the state $|\psi_i\rangle$ results in:

$$cU|\psi_i\rangle = \left(|0\rangle\langle0| \otimes I + |1\rangle\langle1| \otimes \sum_i e^{i\lambda_i} |\lambda_i\rangle\langle\lambda_i|\right) \left(\frac{1}{\sqrt{2}} (|0\rangle + |1\rangle) \otimes |\lambda_i\rangle\right) \tag{29}$$

Simplifying this, we get

$$cU|\psi_i\rangle = \frac{1}{\sqrt{2}} \left(|0\rangle + e^{i\lambda_i}|1\rangle\right) \otimes |\lambda_i\rangle$$

Notice that the target qubit register remains unchanged in the state $|\lambda_i\rangle$, while the control qubit accumulates a phase of $e^{i\lambda_i}$. This phenomenon is known as phase kickback and is widely used in quantum subroutines, such as quantum phase estimation.

### 4.1.2.   Quantum Fourier Transform

The Fourier transform is routinely used in classical scientific computation, implemented as the discrete Fourier transform. Since the Fourier transform is essentially a change of basis, it can be represented as a unitary operation, which can be performed on a quantum computer with exponentially improved scaling.

The Quantum Fourier Transform (QFT) is the quantum analog of the discrete Fourier transform applied to the probability amplitudes of a quantum state [64].

The Fourier transform is exponentially faster on a quantum computer in both resources and time, requiring $O(n^2)$ gates and $n$ qubits compared to the classical complexity of $O(2^n)$ memory requirement and $O(n2^n)$ time complexity.

A discrete Fourier transform takes as input $\boldsymbol{\psi} \in \mathbb{C}^N$ and outputs $\boldsymbol{\phi} \in \mathbb{C}^N$ s.t.

$$\phi_k = \frac{1}{\sqrt{N}} \sum_{j=0}^{N-1} \psi_j e^{2\pi ijk/N}. \tag{30}$$

The quantum Fourier transform maps the basis states $|j\rangle$ of an input state $|\psi\rangle = \sum_{j=0}^{N-1} \psi_j |j\rangle$ to an output state $|\phi\rangle = \sum_{k=0}^{N-1} \phi_k |k\rangle$ as:

$$|j\rangle \rightarrow \frac{1}{\sqrt{N}} \sum_{k=0}^{N-1} e^{2\pi ijk} |k\rangle \tag{31}$$

corresponding to the operation on an arbitrary state $|\psi\rangle$.

$$\sum_{j=0}^{N-1} \psi_j |j\rangle \rightarrow \frac{1}{\sqrt{N}} \sum_{k=0}^{N-1} \sum_{j=0}^{N-1} \psi_j e^{2\pi ijk/N} |k\rangle \tag{32}$$

The probability amplitudes $\phi_k$ of $|\phi\rangle = \sum_{k=0}^{N-1} \phi_k |k\rangle$ are the discrete Fourier transform of the probability amplitudes $\psi_j$ of $|\psi\rangle = \sum_{j=0}^{N-1} \psi_j |j\rangle$.

Another convenient representation of the quantum Fourier transform is its product form. To do this, it is convenient to use the following labeling for the basis states instead:

For $N = 2^n$, where $n \in \mathbb{Z}^+$, the basis states $|j\rangle \in \{|0\rangle, ..., |2^n - 1\rangle\}$ may be relabeled using the binary notation of the integers $j$ as follows

$$j = j_1 2^{-1} + j_2 2^{n-2} + ... + j_n 2^0 : \rightarrow j_1 j_2 ... j_n \tag{33}$$

Similarly, the binary fraction is denoted as:

$$j_l / 2 + j_l + 1/4 + ... + j_m / 2^{m-l+1} : \rightarrow 0. j_l j_{l+1} ... j_m \tag{34}$$

Using the notation, the quantum Fourier transform on a basis state $|j_1 j_2 ... j_n\rangle$ may be rewritten as

$$|j_1 j_2 ... j_n\rangle \rightarrow \frac{(|0\rangle + e^{0.j_n}|1\rangle)(|0\rangle + e^{0.j_{n-1} j_n}|1\rangle)..|0\rangle) + e^{0.j_1 j_2 ... j_n}|1\rangle)}{2^{n/2}} \tag{35}$$

We provide as an example a Qiskit [23] implementation of a quantum Fourier transform on a uniform superposition of basis states below. A uniform superposition of qubits corresponds to an amplitude encoding of a vector of 1's, i.e., $|\Phi\rangle = \frac{1}{\sqrt{N}} \sum_{k=0}^{2^n - 1} |k\rangle$, shown as the "before" state in Figure 12, which corresponds to the $0^{\text{th}}$ frequency component in the Fourier basis. Performing a quantum Fourier transform on $|\Phi\rangle$ yields a quantum state $|\Psi\rangle$, an amplitude encoding of the Fourier transform of the probability amplitudes of $|\Phi\rangle$. Measuring the qubits after the quantum Fourier transform on $|\Phi\rangle$ yields the quantum state $|\Psi\rangle = |00...0\rangle$, the basis state corresponding to the $0^{\text{th}}$ frequency component, as expected shown in Figure 12 as the "after" state.

```python
#!/usr/bin/python3

from matplotlib import pyplot as plt
from qiskit import QuantumCircuit
from qiskit.circuit.library import QFT
from qiskit_aer import Aer
from qiskit.visualization import plot_histogram

beforeFT = QuantumCircuit(5)
# Initialize state in uniform superposition
beforeFT.h([0,1,2,3,4])

# Measure all qubits
beforeFT.measure_all()

# Simulate the circuit
mySimulator = Aer.get_backend('aer_simulator')
result = mySimulator.run(beforeFT,shots=2**20).result()

# Plot a bar chart of all the results
plot_histogram(result.get_counts(),bar_labels=False,title='Before QFT')

afterFT = QuantumCircuit(5)
# Initialize qubits
afterFT.h([0,1,2,3,4])

# Add Fourier transform operation
qft = QFT(num_qubits=5,do_swaps=False).to_gate()
afterFT.append(qft, qargs=[0,1,2,3,4])

# Measure all qubits
afterFT.measure_all()

# Decompose Fourier transform operation into gates for simulator
afterFT = afterFT.decompose(reps=2)
```

```
# Simulate the circuit
result = mySimulator.run(afterFT,shots=2**20).result()

# Plot a bar chart of all the results
plot_histogram(result.get_counts(),bar_labels=False,title='After QFT')
plt.show()
```

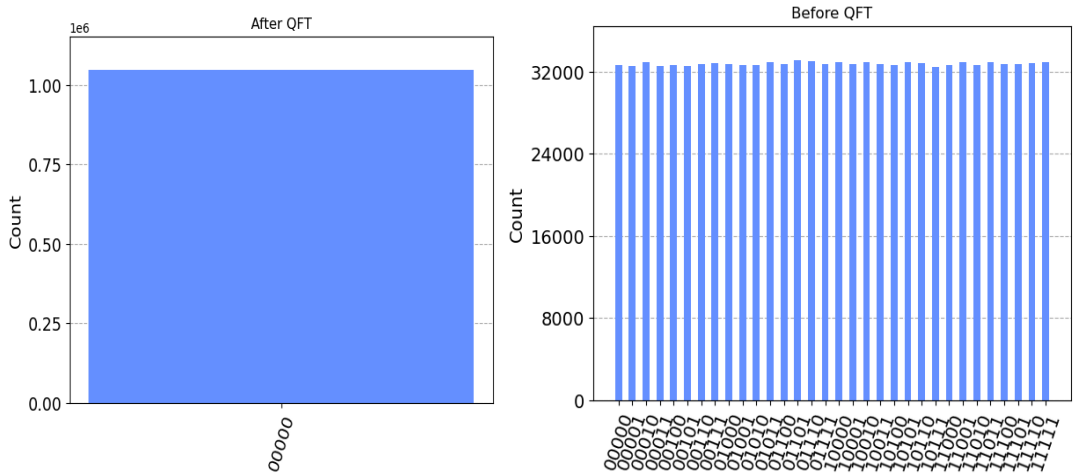

*Figure 12 Basis states before and after a quantum Fourier transform*

### 4.1.3.    Phase Estimation

The phase kickback subroutine effectively provides access to the eigenvalue of a unitary operator by embedding it into the phase of the control qubit. It may be of interest to read out this eigenvalue or use it in further computation with arbitrary precision. The phase estimation subroutine combines the phase kickback and quantum Fourier transform operations to enable this task.

Given an eigenstate $|\lambda_i\rangle$ of a unitary $U$, quantum phase estimation provides an estimate of the corresponding eigenvalue $e^{i\lambda_i}$. Phase estimation uses the quantum Fourier transform to estimate the phase kicked back by a unitary operation. A series of controlled versions of the unitary operation are performed on an input state. Figure 13 shows an example with 4 control qubits. The control qubits are put in uniform superposition before applying the controlled operations. More specifically, $j$ control qubits are used to apply the unitary $U^{2^j}$ controlled by the $j^{th}$ control qubit. Denoting $c_j U^{2^j}$ as the unitary $U^{2^j}$ applied to the work register, controlled by the $j^{th}$ qubit, from the phase kickback result we have:

$$\prod_{j=0}^{J-1}\left(c_jU^{2^j}\right)\left(\frac{1}{2^{\frac{J-1}{2}}}(|0\rangle+|1\rangle)^{\otimes J-1}|\psi\rangle\right) \quad (36)$$

$$=\frac{1}{2^{\frac{J-1}{2}}}\bigotimes_{j=0}^{J-1}(|0\rangle+e^{i\lambda_i2^j}|1\rangle)|\psi\rangle \quad (37)$$

$$=\frac{1}{2^{\frac{J-1}{2}}}\sum_{k=0}^{2^{J-1}-1}(|0\rangle+e^{ik\lambda_i}|1\rangle)|\psi\rangle. \quad (38)$$

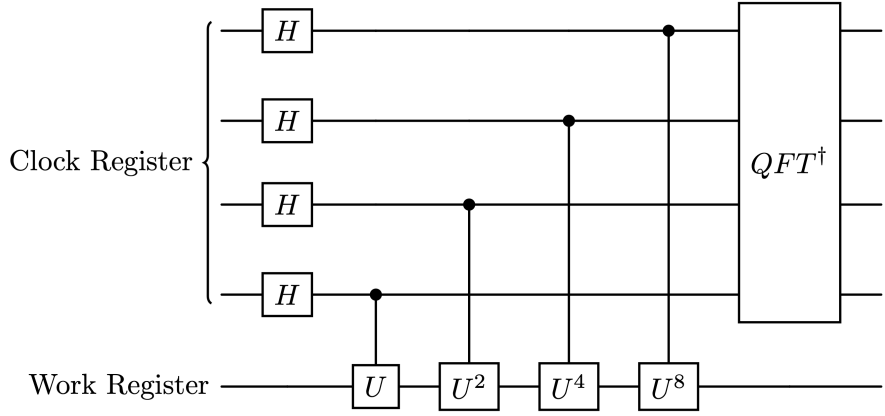

*Figure 13 Circuit for quantum phase estimation.*

The phase estimation algorithm requires $j$ ancilla qubits. A more efficient implementation is the iterative phase estimation algorithm, which requires a single ancilla qubit to perform phase estimation [65] and dynamic circuits.

### 4.1.4.    Amplitude Amplification

Quantum algorithms are inherently probabilistic, and may have a non-zero probability of failure. In general, the success probability can be boosted using the amplitude amplification procedure. Amplitude amplification is an important subroutine used to boost the success probability of the desired output of a quantum algorithm $\mathcal{A}$. The procedure is proven to be optimal for general $\mathcal{A}$ and provides a quadratic improvement in overall complexity [40].

Consider a quantum algorithm $\mathcal{A}$ s.t.

$$\mathcal{A}|0\rangle^{\otimes n}=\sqrt{p}|\psi_{good}\rangle+\sqrt{1-p}|\psi_{bad}\rangle \quad (39)$$

where $|\psi_{good}\rangle$ is the desired output with success probability $p$. The amplitude amplification procedure will boost the success probability to $O(1)$ using $O(1/\sqrt{p})$ applications of $\mathcal{A}$.

More specifically, this is achieved using $O(\sqrt{p})$ applications of the operator $Q$ combining two reflection operations:

$$Q = -\mathcal{S}_0 \mathcal{S}_{good} \tag{40}$$

where

$$\mathcal{S}_0 = \mathbb{I} - 2|0\rangle^{\otimes n}\langle 0|^{\otimes n} \tag{41}$$

is a reflection operation about the state $|0\rangle^{\otimes n}$ and

$$\mathcal{S}_{good} = \mathbb{I} - 2|\psi_{good}\rangle\langle\psi_{good}| \tag{42}$$

is an oracle for reflection about the state $|\psi_{good}\rangle$. Intuitively, these operations can also be interpreted as "marking" the states $|\psi_{good}\rangle$ and $|0\rangle^{\otimes n}$ with a negative phase. Note that the operations are unitary Householder reflectors. [66] provides a guide to implementing the operator $\mathcal{S}_{good}$ given $\mathcal{A}$. The overall circuit for amplitude amplification is provided in Figure 14.

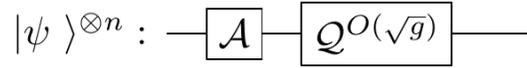

*Figure 14 Circuit for amplitude amplification.*

More advanced algorithms exist for amplitude amplification for specific problems, a notable example being Variable-Time Amplitude Amplification [67], which can provide speedups for algorithms whose average stopping time is smaller than the maximum stopping time by recasting it as a variable-time stopping algorithm. A more efficient version of the Variable-Time Amplitude Amplification sub-routine was presented by [25].

Grover's search algorithm [68] is a special case of amplitude amplification, which searches for an entry in an unstructured database of $N$ entries using $O(\sqrt{N})$ queries to the database compared to $O(N)$ classical queries.

### 4.1.5. Linear Combination of Unitaries

Matrix-vector multiplication is a fundamental task in scientific computing. Since quantum gates are unitary, naïve implementation of operations using quantum gates limits the range of computations to unitary operations. The linear combination of unitaries (LCU) subroutine provides a more general framework for arbitrary matrix-vector multiplications.

The linear combination of unitaries method is a useful algorithmic primitive to apply a linear combination of individual unitaries, which can be non-unitary, to a quantum state [69].

Given oracles or quantum circuits for the unitaries $U_l$ and the coefficients $\alpha_l$ of the linear combination of the unitaries $U = \sum_l \alpha_l U_l$ can be implemented using $SELECT$ and $PREPARE$ operations.

The $PREPARE$ operation implements the following operation:

$$PREPARE|0\rangle^{\otimes m} = |\alpha\rangle \tag{43}$$

where $|\alpha\rangle = \frac{1}{\|\alpha\|_1} \sum_l \sqrt{\alpha_l} |l\rangle$ s.t. $\alpha_l \in \mathbb{R}^+$ and $\|\alpha\|_1$ is the overall subnormalization constant. This is without any loss of generality since any phase of the coefficients can be absorbed into the unitary $U_l$.

The $SELECT$ operation is a sequence of controlled versions of the unitaries $U_l$:

$$\prod_l |l\rangle\langle l| U_l. \tag{44}$$

To apply a linear combination of unitaries, we apply the operations:

$$(PREPARE^\dagger \otimes I)(SELECT)(PREPARE \otimes I). \tag{45}$$

Measuring the ancilla register in the state $|0\rangle^{\otimes m}$ indicates the successful application of the linear combination of unitaries.

Note that the linear combination of unitaries technique is, in fact, a block-encoding of an arbitrary matrix $\boldsymbol{H}$, where $\boldsymbol{H}$ can be non-Hermitian. Block-encoded access to a matrix unlocks the usage of qubitization methods, of which quantum signal processing is a commonly used efficient (in the number of ancilla qubits) version and is introduced in Section 1.3.1.7.

We provide as an example a Qiskit implementation of a circuit taking a linear combination of the following unitaries below:

$$I \otimes I + X \otimes I + I \otimes X + I \otimes Z = \begin{pmatrix} 2 & 1 & 1 & \\ 1 & & & 1 \\ 1 & & 2 & 1 \\ & 1 & 1 & \end{pmatrix} \tag{46}$$

where the coefficients $\alpha_l = 1/4$, for which the corresponding $PREPARE$ operation is simply a uniform superposition achieved by Hadamard gates. The circuit for the example is shown in Figure 15.

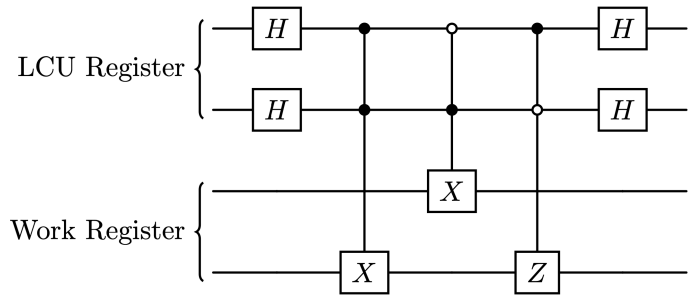

*Figure 15 A circuit implementing a linear combination of unitaries of Pauli matrices.*

```
#!/usr/bin/python3

import qiskit
from qiskit_aer import Aer
from qiskit.circuit.library import XGate, ZGate
import numpy as np

# LCU register
lcuRegister = qiskit.QuantumRegister(2, 'LCU')
# Work register
workRegister = qiskit.QuantumRegister(2, '\psi')

myCircuit = qiskit.QuantumCircuit(workRegister,lcuRegister)

# PREP+ operation
myCircuit.h(lcuRegister)

# SELECT operation
myCircuit.append(XGate().control(num_ctrl_qubits=2,ctrl_state='11'),[*lcuRegi
ster, workRegister[0]])
myCircuit.append(XGate().control(num_ctrl_qubits=2,ctrl_state='01'),[*lcuRegi
ster, workRegister[1]])
myCircuit.append(ZGate().control(num_ctrl_qubits=2,ctrl_state='10'),[*lcuRegi
ster, workRegister[0]])

# PREP operation
myCircuit.h(lcuRegister)

# Simulate circuit
backend = Aer.get_backend('unitary_simulator')
result = backend.run(myCircuit,shots=0).result()
# Extract subspace of successfully measuring LCU qubits as 0
# and multiply by submormalization factor 4
print((np.array(result.get_unitary().data[:4,:4]).round(10)*4)[::-1,::-1])
```

### 4.1.6. Qubitization and Quantum Signal Processing

Beyond matrix-vector multiplications, applying matrix polynomials is a powerful tool in scientific computing. Using block-encoded or "qubitized" matrices, the quantum signal processing method allows the application of matrix polynomials to quantum statevectors.

Qubitization is a powerful method in quantum computing that creates an $SU(2)$ subspace for each eigenvalue of a matrix [70], which can then be used to generate polynomials of the eigenvalues. Consider a $2 \times 2$ rotation matrix:

$$O(\lambda) = \begin{pmatrix} \lambda & -\sqrt{1-\lambda^2} \\ \sqrt{1-\lambda^2} & \lambda \end{pmatrix}. \tag{47}$$

By applying powers of this matrix, we get:

$$O^k(\lambda) = \begin{pmatrix} T_k(\lambda) & -\sqrt{1-\lambda^2}U_{k-1}(\lambda) \\ \sqrt{1-\lambda^2}U_{k-1}(\lambda) & T_k(\lambda) \end{pmatrix} \tag{48}$$

where $T_k(\lambda)$ and $U_k(\lambda)$ are Chebyshev polynomials of the first and second kind, respectively. If the scalar $\lambda$ is replaced with a matrix $\boldsymbol{A}$, then the matrix

$$O^k(\boldsymbol{A}) = \begin{pmatrix} T_k(\boldsymbol{A}) & -\sqrt{1-\boldsymbol{A}^2}U_{k-1}(\boldsymbol{A}) \\ \sqrt{1-\boldsymbol{A}^2}U_{k-1}(\boldsymbol{A}) & T_k(\boldsymbol{A}) \end{pmatrix} \tag{49}$$

is a block-encoding of $T_k(\boldsymbol{A})$. Using a linear combination of these Chebyshev polynomials (using LCU techniques introduced in Section 4.1.5), any arbitrary polynomial of the matrix $\boldsymbol{A}$ can be created.

The goal of qubitization is to use any arbitrary block-encoding of the form

$$U_A = \begin{pmatrix} \boldsymbol{A} & * \\ * & * \end{pmatrix} \tag{50}$$

to generate polynomials $P(\boldsymbol{A})$ of the matrix $\boldsymbol{A}$. This is achieved by using an ancilla qubit to apply the operation $O_A = U_A Z_\Pi$, which is the desired equivalent of $O(\boldsymbol{A})$, as shown in the circuit in Figure 16.

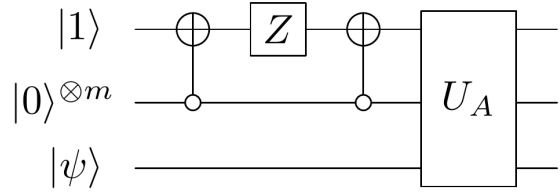

*Figure 16: The operation $\mathbf{Z_\Pi}$ followed by $\mathbf{U_A}$.*

For a Hermitian $\mathbf{A}$, repeated application of $O_A = U_A Z_\Pi$ will yield the desired Chebyshev polynomial of $\mathbf{A}$. For non-Hermitian $\mathbf{A}$, the alternating sequence $O_A = U_A Z_\Pi U_A^\dagger Z_\Pi$ can be used instead.

Since the block-encoding of the Chebyshev polynomial (an orthogonal polynomial basis) is a unitary operation, any arbitrary polynomial can be constructed using a linear combination of these unitaries. However, this would require an overhead of up to $O(\log(d))$ additional qubits where $d$ is the degree of the polynomial.

Quantum signal processing (QSP) [71] addresses the issue of additional overheads by forming a polynomial directly by applying rotation operations to the ancilla qubit. The main idea behind quantum signal processing is to replace the $Z$ gate in qubitization with the general $Z$ rotations $R_Z(2\phi_i)$, where $\phi_i \in \Phi \in \mathbb{R}^{d+1}$ is a predetermined sequence of $d+1$ rotation angles to form a block-encoding of the desired polynomial $P(\mathbf{A})$:

$$P(\mathbf{A}) = \langle 0|^{\otimes m} U_\Phi(\mathbf{A})|0\rangle^{\otimes m} \tag{51}$$

$$U_\Phi(\mathbf{A}) = e^{-i\phi_0 Z} \prod_{j=1}^{d} \left[ U_A e^{-i\phi_j Z} \right] \tag{52}$$

where $\mathbf{A}$ is Hermitian. An overview of a quantum signal processing circuit for a non-Hermitian block-encoding $U_A$ of a matrix $\mathbf{A}$ is given in Figure 17, where block-encoding and its Hermitian conjugate are used in an alternating sequence. The quantum signal processing theorem [26] states that a sequence of $d+1$ rotation angles exists for any complex polynomial $P$ of maximum degree $d$, and the polynomial is even for even $d$ and odd for odd $d$. Furthermore, it states that the complex conjugate of the polynomial can be formed using the phase angles $-\Phi$.

**Theorem 7** (Quantum Signal Processing): [26] The quantum signal processing sequence $U_\Phi$ produces a matrix that may be expressed as a polynomial function of x:

$$U_\Phi = e^{i\phi_0 Z} \prod_{k=1}^{d} W(x) e^{i\phi_k Z} = \begin{pmatrix} P(a) & iQ(a)\sqrt{1-a^2} \\ iQ^*(a)\sqrt{1-a^2} & P^*(a) \end{pmatrix} \qquad (53)$$

For $x \in [-1,1]$, and a $\Phi$ exists for any polynomials $P, Q$ in $x$ s.t.:

i) $\text{Deg}(P) \le d, \deg(Q) \le d - 1$ \hfill (54)

ii) $P$ has a parity (is even or odd) $d \bmod 2$ and $Q$ has a parity $(d - 1) \bmod 2$

iii) $|P|^2 + (1 - a^2)|Q^2| = 1$.

Using this result, we note that any arbitrary polynomial can be formed by taking a linear combination of the even and odd parts of the polynomial. Furthermore, the real or complex parts of a polynomial can be extracted by taking a linear combination of the polynomial and its complex conjugate. There are various libraries for calculating the required phase angles for any desired polynomial [26], [60], with QSPPACK [60] providing state-of-the-art performance at the time of writing this paper.

Recent work dubbed Generalized Quantum Signal Processing has generalized the Z rotations on the ancilla qubit to include X and Y rotations [72], which allows arbitrary polynomials of degree $d$ to be constructed without taking linear combinations of the even and odd parts, while also making the calculation of the phase angles efficient and accurate, albeit it requires access to a block-encoding of $e^{iA}$ instead of $A$.

Quantum signal processing is currently the most powerful and optimal method for many quantum algorithms. The quantum eigenvalue transform, quantum singular value transform [32], factoring, phase estimation, Hamiltonian simulation, linear system solution, amplitude amplification, eigenstate filtering, and many other quantum computing problems can be reformulated as a quantum signal processing problem with optimal or near-optimal scaling results [26].

In Figure 17, we show the implementation of a controlled rotation using a phase angle $\phi_i$ and a quantum signal processing circuit implementing the sequence of controlled rotations $\Phi$ to implement a polynomial of a matrix $\boldsymbol{A}$ using its block-encoding $U_A$.

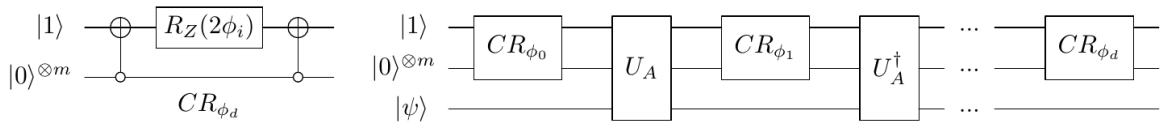

*Figure 17 The controlled rotation operation $\boldsymbol{CR_{\phi_d}}$ and a quantum signal processing circuit for a*

# 5. Quantum Algorithms for Engineering

In this Section, we provide an overview of important algorithms used in quantum computers. Quantum algorithms for scientific and engineering computation need a significant rethinking due to fundamental differences between classical computing (Turing machine) models and the gate-based quantum computing model. Algorithms for classical computing have been refined over the years and have optimized libraries (e.g. BLAS for linear algebra computations) and dedicated hardware components (Floating Point Units in processors and GPU accelerators) for scientific computing. However, quantum computing is in its nascent stages. Current research in quantum computing is driven by hardware error rates and algorithm development.

In Section 5.1 we provide introduce the Hamiltonian simulation problem, a core problem in quantum scientific computing, and various approaches for its approximate solution on quantum computers.

We then define a quantum linear system problem in Section 5.2 and provide an overview of three major quantum linear system algorithms, which are analogous to classical direct solvers, in Sections 5.2.1-5.2.3. Iterative methods for linear systems and preconditioners, which are an active area of research, are discussed in Section 5.2.4.

In Section 5.3 algorithms for various classes of ordinary differential equations are covered. We then discuss a variety of prominent quantum algorithms for partial differential equations in Section 5.4.

## 5.1. Hamiltonian Simulation

Hamiltonian simulation is the problem of applying the operation $U \approx e^{-iHt}$ to a quantum state. When $\boldsymbol{H}$ is a Hermitian matrix, $U$ is unitary. Hamiltonian simulation is important for simulating closed quantum systems (quantum systems which do not interact with the environment or external fields) like Hubbard models for condensed matter physics. It is also used as a subroutine in various quantum algorithms like the HHL quantum linear system algorithm [5]. Hamiltonian simulation encodes the evolution of a quantum state according to the time-dependent Schrodinger equation:

$$\frac{d}{dt}|\Psi(t)\rangle = -i\boldsymbol{H}(t)|\Psi(t)\rangle \tag{55}$$

where the Planck constant has been absorbed into the Hamiltonian $\boldsymbol{H}(t)$. In this review, we consider the case of time-independent Hamiltonians, i.e. $\boldsymbol{H} \neq \boldsymbol{H}(t)$, which has the analytical solution

$$|\Psi(t)\rangle = e^{-i\boldsymbol{H}t}|\Psi(0)\rangle. \tag{56}$$

One can easily see that the Hamiltonian simulation of a closed quantum system is simply the solution of a homogeneous first-order system of ordinary differential equations. Various methods exist for Hamiltonian simulation, the most important ones being product formulas, Taylor series approximations using linear combinations of unitaries, and quantum signal processing. A plethora of other methods have been developed based on randomized evolutions [73], [74]. In Section 5.1.1 we introduce the 1st- and 2nd-order Suzuki-Trotter product formulae and analyze their approximation errors with an implementation in Qiskit. We then introduce the Taylor series approach for Hamiltonian simulation in Sections 5.1.2 and 5.1.3, which scales exponentially better in approximation errors. The quantum signal processing approach, which scales optimally, is presented in Section 5.1.3.

We discuss the solution of more general systems of ordinary differential equations in more detail in Section 5.3, where techniques like time-marching strategies are presented.

Hamiltonian simulation has great potential for speeding up scientific and engineering computation. All homogeneous linear systems of differential equations can be transformed into a system of first-order differential equations, which can then be solved using Hamiltonian simulation techniques. The speedups for quantum chemistry itself can be immense, allowing *ab-initio* computation of larger systems with higher accuracy. Density functional theory (DFT) is a commonly used method for *ab-initio* calculations. However, DFT methods typically rely on approximations and usually can only provide ground-state solutions, which limits their application. Direct estimation of observables from a simulation of the quantum state can provide more accurate properties for multiscale modeling and can unlock interesting physics coming from the excited states of molecules and crystals.

We note that Hamiltonian simulation for various problems requires a preprocessing step to map to a quantum computer. As an example, for quantum chemistry problems, the Jordan-Wigner [58] or Bravyi-Kitaev [59] transforms can be used to map the second quantization of an atom or molecule (a Fermionic Hamiltonian) to qubits and unitary operations of quantum computers [75]. To realize the exponential speedups offered by quantum computers for Hamiltonian simulation, efficient implementations of oracles or block-encodings need to be developed for problems of practical interest.

### 5.1.1.    Product Formulas: Suzuki-Trotter

Suzuki-Trotter [76] or Trotter formulas use the Baker-Campbell-Hausdorff formula to approximate

$$e^{-iHt} = e^{-i\sum_{j=0}^{k} H_j t} = \left( e^{-i\sum_{j=0}^{k} H_j \frac{t}{r}} \right)^r \approx \left( \prod_{j=0}^{k} e^{-iH_j \frac{t}{r}} \right)^r + O\left( \frac{k^2 t^2}{r} \right). \tag{57}$$

The decomposition of $\boldsymbol{H}$ is typically done as the sum of 1-sparse $\boldsymbol{H}_j$ (typically Pauli strings) since the matrix exponential $e^{-iH_j t}$ is easily computed and implemented using quantum gates. The error in Trotterization arises from the fact that the matrices $\boldsymbol{H}_j$ do not commute in general, i.e., the commutator $[\boldsymbol{H}_j, \boldsymbol{H}_k] \neq 0$. Higher-order Trotter formulas can be constructed to reduce the error further. Although the error bounds on Trotter formulas do not scale well, Trotter methods have the advantage of not requiring additional (ancilla) qubits and the circuit implementations of the individual $\boldsymbol{H}_j$ can be straightforward. As an example, the 2nd-order Trotter formula is

$$e^{-iHt} = e^{-i\sum_{j=0}^{k} H_j t} = \left( e^{-i\sum_{j=0}^{k} H_j \frac{t}{r}} \right)^r \approx \left( \prod_{j=0}^{k} e^{-iH_j \frac{t}{r}} \prod_{j=k}^{0} e^{-iH_j \frac{t}{r}} \right)^r + O\left( \frac{k^3 t^3}{r^2} \right). \tag{58}$$

Typically, $\boldsymbol{H}$ is implemented as a sum of Pauli strings $\boldsymbol{H}_j$. Therefore, the $e^{-iH_j \frac{t}{r}}$ terms can be implemented trivially as single-qubit rotation gates for Pauli matrices $\boldsymbol{H}_j \in \{X, Y, Z\}$. Procedures to construct quantum circuits exponentiating Pauli strings with more than one non-trivial term, e.g. $I \otimes I \otimes X \otimes Y \otimes Z \otimes I \otimes I$, can be found in [1], [77].

As shown in Equation 58, Trotter errors scale as $O(\text{poly}(1/\epsilon))$ and can be improved by either increasing the number of Trotter steps or using higher order Trotter formulae. Recent work has shown higher-order formulae with improved polynomial scaling of errors compared to Trotter formulae [78]. The bounds on Trotter errors are known to be loose, and in practice, the errors have been shown to be orders of magnitude lower [7], [79]. As an example, in Figure 18 the approximation error for the 1ˢᵗ order approximation is unexpectedly lower for $t = 2$ compared to $t = 1$ for the same number of Trotter steps. Methods with logarithmic scaling in errors have been developed and are presented in subsequent sections.

We provide in Appendix A2 an example code for a Hamiltonian simulation of $\boldsymbol{H} = X + Z$ using the 1ˢᵗ - and 2ⁿᵈ- order Trotter method for various $t$, for varying numbers of Trotter steps, with the error plots shown in Figure 18.

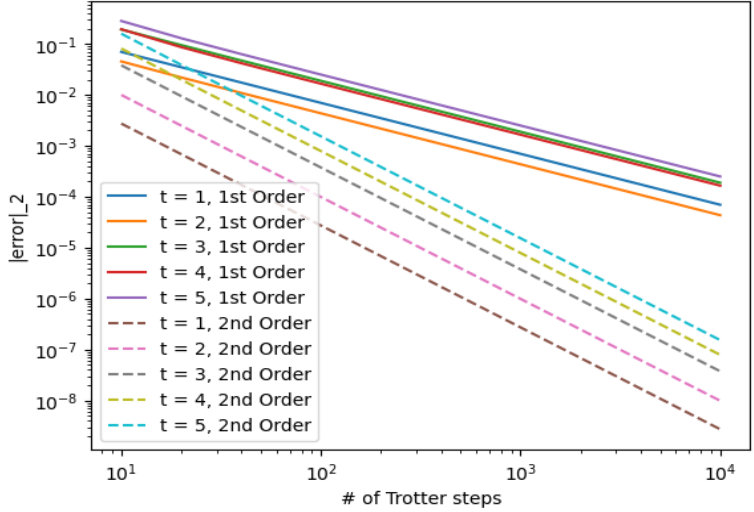

*Figure 18 Error scaling of 1ˢᵗ and 2ⁿᵈ order Suzuki-Trotter methods*

### 5.1.2.    Taylor Series

Since $U = e^{-i\boldsymbol{H}t}$ is an analytical function, it may be approximated using a truncated Taylor series. The Taylor series approximation for Hamiltonian simulation is derived by decomposing $\boldsymbol{H}$ as a sum of unitaries $\boldsymbol{H}_l$ [80] similar to the product formula approach

$$U = e^{-i\boldsymbol{H}t} = (e^{-i\boldsymbol{H}t/r})^r = (U_r)^r \tag{59}$$

$$U_r \approx \widehat{U}_r = \sum_{k=0}^{K} \sum_{l_1,\dots,l_k=1}^{L} \frac{(-it/r)^k}{k!} \alpha_{l_1} \dots \alpha_{l_k} \boldsymbol{H}_{l_1} \dots \boldsymbol{H}_{l_k}. \tag{60}$$

A choice of $K = O\left(\frac{\log\left(\frac{r}{\epsilon}\right)}{\log\log\left(\frac{r}{\epsilon}\right)}\right)$ leads to a precision $\parallel U - (\widehat{U}_r)^r \parallel_2 \leq \epsilon$, an exponential improvement over Trotter simulations.

If $\boldsymbol{H}_l$ are Pauli strings, their products $\boldsymbol{H}_{l_1} \dots \boldsymbol{H}_{l_k}$ are also unitary. Therefore, the double summation may be implemented using the linear combination of unitaries subroutine, which requires ancilla qubits and has a non-zero probability of failure. We note that if the LCU subroutine is implementing (approximately) unitary operations its success probability can be boosted using to $O(1)$ using a subroutine known as Oblivious Amplitude Amplification [34]. The overall algorithm simulates a $d$-sparse Hamiltonian $\boldsymbol{H}$ for time $t$ to a precision of $\epsilon$ with $O\left(\tau \frac{\log^2 \tau}{\epsilon} \frac{1}{\log\log\left(\frac{\tau}{\epsilon}\right)}\right)$ queries to $\boldsymbol{H}$, where $\tau = d^2 \|\boldsymbol{H}\|_{max} t$, which is near optimal in time.

This approach requires increasing numbers of ancilla qubits and a large number of controlled operations. The Quantum signal processing method, presented in the next section, requires significantly fewer ancillae and controlled operations. Nevertheless, these approaches are not amenable to non-fault-tolerant quantum computers.

### 5.1.3.    Quantum Signal Processing

The quantum signal processing approach for Hamiltonian simulation uses Euler's formula to split the exponential operator:

$$e^{-i\boldsymbol{H}t} = \cos(-i\boldsymbol{H}t) + \sin(-i\boldsymbol{H}t) = \cos(i\boldsymbol{H}t) - \sin(i\boldsymbol{H}t) \tag{61}$$

and implements $\cos(i\boldsymbol{H}t)$ and $\sin(i\boldsymbol{H}t)$ separately using the quantum signal processing method and uses the LCU approach to implement their sum. The terms $\cos(i\boldsymbol{H}t)$ and $\sin(i\boldsymbol{H}t)$ are $\epsilon$-approximated using the Jacobi-Anger expansion:

$$\cos(xt) \approx P_{\cos}(x) = J_0(t) + 2\sum_{k=1}^{k'} (-1)^k J_{2k}(t) T_{2k}(x) \tag{62}$$

$$\sin(xt) \approx P_{\sin}(x) = 2\sum_{k=0}^{k'} (-1)^k J_{2k+1}(t) T_{2k+1}(x) \tag{63}$$

where $J_i(x)$ is the $i^{th}$ order Bessel function and $T_i(x)$ is the $i^{th}$ Chebyshev polynomial of the 1$^{st}$ kind with a choice of $k' = \left\lfloor \frac{1}{2} r\left(\frac{e}{2}|t|, \frac{5}{4}\epsilon\right)\right\rfloor$.

Implementing these polynomials using QSP requires choosing a truncation error of $\epsilon/4$ and rescaling and the polynomials by a factor of $\frac{1}{1+\epsilon/4}$ to ensure that both $\|P_{\cos}(x) + P_{\sin}(x)\| \leq 1$ and $\|P_{\cos}(x) + P_{\sin}(x) - e^{ix}\| \leq \epsilon$. This approach achieves optimal scaling in $t$ by simulating a $d$-sparse Hamiltonian for time $t$ with error $\epsilon$ using $O(td\|\boldsymbol{H}\|_{max} + \frac{\log(1/\epsilon)}{\log\log(1/\epsilon)})$ queries to $\boldsymbol{H}$ [71]. Note that since the approximations of $\cos(i\boldsymbol{H}t)$ and $\sin(i\boldsymbol{H}t)$ are implemented using QSP for $t \geq 0$, this approach only holds for positive-definite $\boldsymbol{H}$. If $\boldsymbol{H}$ is not positive-definite, this can be remedied by using the block-encoding of $\boldsymbol{H}$ to implement a block-encoding $\boldsymbol{H}_{+} = \frac{1}{2}(\boldsymbol{H}/\alpha + \boldsymbol{I})$ instead where $\alpha$ is the subnormalization factor of the block-encoding of $\boldsymbol{H}$. $\boldsymbol{H}_{+}$ is positive definite and $e^{-2i\boldsymbol{H}_{+}\alpha t}$ is equivalent to $e^{-i\boldsymbol{H}t}$ up to a global phase factor.

## 5.2. Quantum Linear System Algorithms

Quantum linear system algorithms (QLSA) hold critical importance for quantum computing in scientific and engineering calculations, given the prevalence of linear systems across these fields. Classical solvers for linear systems, in general, require a time proportional to the system size to write out the solution of a linear system, leading to a lower bound of $O(N)$ complexity. However, practically, the entire solution is typically not of interest; rather, some scalar properties of a solution are sought. As an example, one may be interested in the maximum stress in a structure under load or the lift or drag produced by the pressure distribution over an airfoil.

To address this issue, the QLSAs solve the quantum linear system problem (QLSP) [5], which is slightly different from the classical linear system problem and can be stated as:

**Definition 6**: (Quantum Linear System Problem) Given a quantum state $|\boldsymbol{b}\rangle$ and access to the matrix $\boldsymbol{A}$, prepare a state $|\boldsymbol{x}\rangle$ with precision $\||\tilde{\boldsymbol{x}}\rangle - |\boldsymbol{x}\rangle\|_2 \leq \epsilon$ s.t. $\boldsymbol{A}|\tilde{\boldsymbol{x}}\rangle = |\boldsymbol{b}\rangle$.

Due to the normalization of quantum states, the quantum state $|\boldsymbol{x}\rangle$ is proportional to the solution $\boldsymbol{x}$. Furthermore, unlike classical solutions, the probability amplitudes of the quantum state $|\boldsymbol{x}\rangle$ encode the solution instead of the entire solution being available classically.

Since the entire quantum state can be operated on efficiently, this enables exponential speedups in $N$ for QLSPs [5]. Multiple QLSAs have been developed, of which the most notable are the HHL algorithm [5], the LCU algorithm [4], and the QSP algorithm [32]. A QLSA with optimal asymptotic scaling was proposed by [81]. We summarize various QLSAs in Table 2.

*Table 2 A summary of notable QLSAs.*

| Algorithm | Complexity | Pros | Cons | Notes |
|-----------|-----------|------|------|-------|
| HHL [5] (2007) | $O(d^2 \kappa^2 \log(N)/\epsilon)$ | 1 ancilla Short circuit possible | High error | $\kappa^2 \to \kappa \log \kappa$ using VTAA [67] |
| LCU [4] (2017) | $O(\kappa^2 \operatorname{poly} \log(\kappa N/\epsilon))$ | Many ancillae | Low error Complex circuit | $\kappa^2 \to \kappa \log \kappa$ using AA [4] |
| QSP [26] (2021) | $O(\kappa \log(\kappa N/\epsilon))$ | Few ancillae | QSP sequence $\Phi$ required | $\Phi$ can be reused |
| PD-QLSA [82] (2021) | $O(\sqrt{\kappa} \log(\kappa N/\epsilon))$ | $\approx$ optimal $\sqrt{\kappa}$ classical scaling | Only for SPD systems | Requires upper bound on $\|A\|_2$ |
| Adiabatic Evolution [83] (2022) | $O(\kappa \log(N/\epsilon))$ | Optimal $\kappa$ scaling | Requires construction of a Hamiltonian for evolution | |

### 5.2.1.    HHL Algorithm

The Harrow-Hassidim-Lloyd (HHL) algorithm is the first proposed algorithm to demonstrate an exponential advantage in system size for linear systems of equations, scaling as $O(\kappa^2 d^2 log(N)/\epsilon)$ where $d$ is the maximum number of nonzero entries in any row or column of $A$. The HHL algorithm uses Hamiltonian simulation to apply controlled versions of $e^{-iAt}$ on the state $|b\rangle$. Since $A$ and $e^{-iAt}$ share the same eigenvectors, the eigenvalues of $e^{-iAt}$ are kicked back to the control qubits. Using the quantum Fourier transform, eigenvalues are encoded in the amplitudes of the control qubits, and a controlled rotation on an ancilla is used to invert the eigenvalues. The computation is then reversed to disentangle the registers, and the ancilla is measured. Measurement of the ancilla in the desired state indicates a successful solution. Since the algorithm relies on phase estimation, it suffers from poor scaling in precision. Variable-time amplitude amplification can be used to improve the dependence on the condition number to linear scaling in $\kappa$ [67]. An overview of the circuit for the HHL algorithm is given in Figure X.

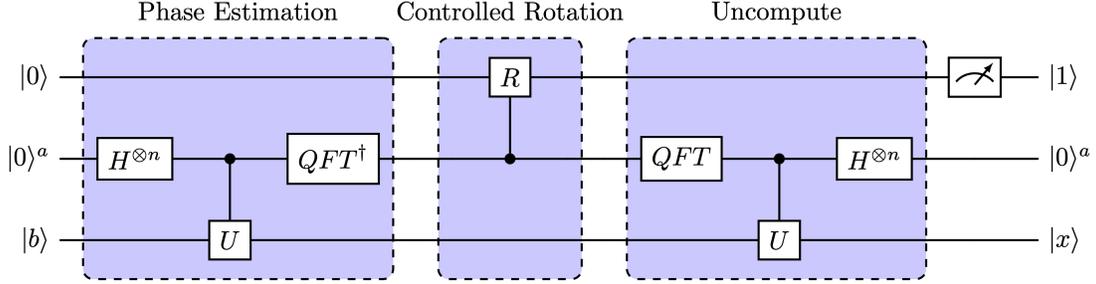

*Figure 19 A quantum circuit demonstrating the HHL algorithm.*

## 5.2.2. LCU QLSA

The LCU algorithm [4] uses instead either a Chebyshev polynomial approximation $P(x) = \sum_k T_k(x) \approx 1/x$, where $T_k$ are Chebyshev polynomials of the 1st kind, or a Fourier approximation $G(x) = \sum_j e^{it_j x} \approx 1/x$, where $t_j \in \mathbb{R}$, over the interval $x \in I_P : [-1, -1/\kappa] \cup [1/\kappa, 1]$, where $\kappa$ is the condition number of the system. The matrix polynomial approximation $P(\boldsymbol{A})$ or Fourier approximation $G(\boldsymbol{A})$ is applied to the state $|\boldsymbol{b}\rangle$ to obtain the approximate solution $\boldsymbol{A}^{-1}\boldsymbol{b} \approx P(\boldsymbol{A})\boldsymbol{b}$ (or $\boldsymbol{A}^{-1}\boldsymbol{b} \approx G(\boldsymbol{A})\boldsymbol{b}$). Given a desired precision $\left\| P(x) - \frac{1}{x} \right\|_{\max x \in I_P} \leq \epsilon$ or $\left\| G(x) - \frac{1}{x} \right\| \leq \epsilon$, $P(x)$ is a polynomial of degree $O\left(\kappa \log \frac{\kappa}{\epsilon}\right)$ and $G(x)$ is a Fourier expansion with $O\left(\kappa \sqrt{\log \frac{\kappa}{\epsilon}}\right)$ terms. The terms $T_k(x)$ can be formed either using matrix multiplications through qubitization or quantum walks (an equivalent approach for forming Chebyshev polynomials on gate-based quantum computers) and $G(x)$ can be formed using Hamiltonian simulation.

However, the application of a linear combination of unitaries has an additional overhead in the form of ancilla qubits. The overall algorithm requires $O\left(d\kappa \operatorname{poly} \log\left(\frac{d\kappa}{\epsilon}\right)\right)$ queries to a sparse access oracle for $\boldsymbol{A}$ and requires $O\left(d\kappa \operatorname{poly} \log\left(\frac{d\kappa N}{\epsilon}\right)\right)$ resources. Although its implementation is rather involved, the LCU QLSA is a seminal development for QLSAs with exponentially improved scaling in $\epsilon$.

### 5.2.3. QSP QLSA

The QSP method [32] circumvents the issue of ancilla qubits. This entails finding $\Phi$, the quantum signal processing angles corresponding to the desired polynomial and a block-encoding of the matrix, given a bound on the condition number $\kappa$ and the desired precision $\epsilon$. The sequence $\Phi$ is not specific to the problem and can be reused for any other problem as long as $\kappa_{new} \leq \kappa$ and $\epsilon_{new} \leq \epsilon$. Given a block-encoded oracle for $\boldsymbol{A}$, the query complexity of the algorithm is $O\left(\kappa \log\left(\frac{\kappa}{\epsilon}\right)\right)$, and it requires $O\left(\kappa \log\left(\frac{\kappa N}{\epsilon}\right)\right)$ resources.

Optimal scaling in $\kappa$ has been achieved by [81], with a query complexity of $O\left(\kappa \log\left(\frac{1}{\epsilon}\right)\right)$ using the adiabatic theorem.

For symmetric positive-definite systems, it is possible to achieve the $\sqrt{\kappa}$ scaling of classical solvers [82]. This is done by first defining the alternative polynomial approximation $Q(y) \approx y = \frac{1}{1-x}$ over the interval $y \in I_Q : [-1, 1/\kappa]$, and defining $\boldsymbol{B} = \boldsymbol{I} - \eta \boldsymbol{A}$ where $\eta$ is chosen s.t. $\parallel \boldsymbol{B} \parallel \leq 1$. Given a desired precision $\left\| Q(x) - \frac{1}{x} \right\|_{\max x \in I_Q} \leq \epsilon$, $Q(x)$ is a polynomial of degree $O(\sqrt{\kappa} \log(\kappa/\epsilon))$. The matrix polynomial approximation $Q(\boldsymbol{B})$ is then applied to the quantum state $|\boldsymbol{b}\rangle$.

We provide in Appendix A3 an example of a QLSA implementation using quantum signal processing in Qiskit below. Figure 20 shows the agreement of the QSP polynomial with $1/x$ with $\epsilon \leq 10^{-9}$. The phase factors $\Phi$ are obtained from the QSPPACK library.

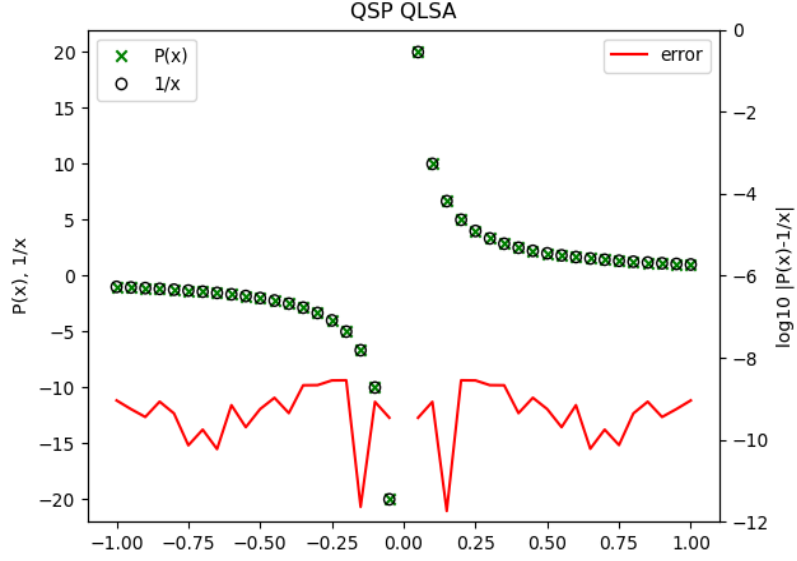

*Figure 20 Output of QSP QLSA code*

### 5.2.4.    Preconditioning and Iterative QLSAs

The linear scaling in $\kappa$ for general linear systems implies that QLSAs can only provide an exponential speedup for problems with $\kappa = O(\text{poly} \log N)$. Practical problems of interest rarely exhibit such scaling. [84] has proposed a quantum sparse approximate inverse (SPAI) preconditioner to resolve this issue. The approach produces the preconditioner by solving a least-squares problem on each row of the matrix. However, the SPAI preconditioner may be inefficient for many problems [85], and a complete implementation of the algorithm is unavailable. A circulant matrix preconditioner has been proposed by [86]. However, its complexity depends on the condition number of the preconditioner $\kappa(\boldsymbol{M})$ and the product $\kappa(\boldsymbol{M^{-1}A})$ which in the worst case can be $\kappa(\boldsymbol{M}) \geq \kappa(\boldsymbol{A})$ and $\kappa(\boldsymbol{M}), \kappa(\boldsymbol{M^{-1}A}) \approx \sqrt{\kappa(\boldsymbol{A})}$. A Laplacian preconditioner has been proposed by [85] for hydrological subsurface flow to reduce the condition number to at most $O(\sqrt{N})$. [87] proposes a fast-inversion procedure to solve QLSPs of the form $(\boldsymbol{A} + \boldsymbol{B})|\boldsymbol{x}\rangle = |\boldsymbol{b}\rangle$ where $\|\boldsymbol{A}\| \ll \|\boldsymbol{B}\|$ with applications towards computing single-particle Green's functions of quantum many-body systems.

Several state-of-the-art quantum algorithms for inhomogeneous ordinary differential equations rely on QLSAs as an intermediate step, which enables their exponential speedups [88], [89].

Numerical solutions of partial differential equations can greatly benefit from QLSAs if the condition number can be managed.

An iterative approach has been proposed by [90] to manage the condition number of the systems of equations using linear stationary iterations, scaling as $O(l \log N/\epsilon)$ (using optimal QLSAs as a subroutine), with possible application to a quantum multigrid method [91] or a quantum domain decomposition method. We discuss quantum algorithms for ODEs and PDEs in Sections 5.3 and 5.4, respectively.

Since the QLSA prepares a quantum state encoding the solution, it also raises the question of computing practically relevant properties of the output. [84] proposed using the QLSA to compute the electromagnetic scattering cross-section of an arbitrary target. [92] have pointed out in their analysis that while the QLSA provides an exponential speedup in preparing the quantum state encoding the solution, using the quantum state to compute properties results in a polynomial speedup for finite element problems when the precision of the output is taken into consideration, with the speedup being greater for higher dimensional problems.

In a subsequent analysis, [93] evaluated the possible quantum speedup for the heat equation by comparing five classical methods with five quantum methods. They show that the direct application of QLSAs to the heat equation is never faster than classical algorithms, but an approach based on amplification and random walks can provide a quadratic speedup for $d \geq 2$, with higher-order speedups for increasing $d$. The heat equation is particularly important since it discretizes the Laplacian operator, which can be used to solve various PDEs as shown by [39].

### 5.3. *Quantum Algorithms for Ordinary Differential Equations*

Quantum algorithms for systems of ordinary differential equations may broadly be classified by homogeneous vs. inhomogeneous, linear vs. nonlinear, and time-dependent vs. time-independent problems. The approaches taken in literature for quantum algorithms for differential equations are to either map the problem to the Schrodinger equation and solve it using Hamiltonian simulation, map the problem to a linear system and solve the linear system using a QLSA, or use the Fourier Transform to diagonalize the system and propagate it in time.

Speedups for ordinary differential equations can have a great impact on engineering and scientific computation. As an example, the $N$-body problem involves solving a system of ordinary differential equations of size $6N$ for time $t$. Compared to a classical complexity of $O(t)$ and $O(N)$ for solution time and memory resources, a quantum computer can potentially solve the problem in $O(t)$ time using $O(\log(N))$ qubits.

The no-fast-forwarding theorem prohibits sublinear scaling in time for general problems [43]. While [39] provides lower bounds on the complexity of solving any general linear systems of ODEs, they also identify specific cases for which the solution can be fast-forwarded, i.e., solved with sublinear scaling in time using a time-marching approach. Bounded negative-semi-definite linear ODE problems with square-root access to $\boldsymbol{A}$ or negative-definite problems can achieve a quadratic speedup in $T$, while problems for which the eigendecomposition is known can achieve an exponential speedup.

Hyperbolic and parabolic partial differential equations, ubiquitous in engineering, physics, finance, medicine, and many other applications, can be transformed into systems of ordinary differential equations using a discretization in space. We discuss partial differential equations in Section 5.4.

### 5.3.1. Homogeneous Linear Systems

The Hamiltonian simulation problem, discussed in detail in Section 5.1, for quantum dynamics of closed quantum systems is solved by the Schrodinger equation, which is a homogeneous system of first-order differential equations. The first quantum algorithm for this problem was proposed by [94] using the Trotter method. The complexity of this method was later improved to a query complexity of $O(d^2 \|\boldsymbol{H}\|_{max} \, t \log(d^2 \|\boldsymbol{H}\|_{max}/\epsilon))$ by [80] by using a linear combination of unitaries arising from a truncated Taylor series. The optimal algorithm for Hamiltonian simulation using quantum signal processing was later provided by [71] with a query complexity of $O(td\|\boldsymbol{H}\|_{max} + \log\left(\frac{1}{\epsilon}\right)/\log\log\left(\frac{1}{\epsilon}\right))$ using $d$-sparse oracles or $O(t\|\boldsymbol{H}\| + + \log\left(\frac{1}{\epsilon}\right)/\log\log\left(\frac{1}{\epsilon}\right))$ using block-encoded oracles.

Recent work has demonstrated methods to circumvent the construction of a large linear system by using a time-marching strategy to propagate the solution in time [95] for homogeneous systems. However, these techniques are sub-optimal and scale quadratically in $t$.

### 5.3.2.  Inhomogeneous Linear Systems

For inhomogeneous linear systems, the first efficient algorithm was provided by [88] by using a linear multistep method for time discretization. The recursive relation for discrete time-stepping is encoded into a larger linear system as back-substitution, which is then solved using a QLSA. The scaling was subsequently improved in [89] to obtain quasilinear scaling in $t$ and exponentially improved precision by using a Taylor series time discretization instead of a linear multistep formula. A spectral approach for time discretization was provided by [96], which can also accommodate the case of time-dependent $A(t)$. We note that all these algorithms require that $A$ have non-positive real parts of its eigenvalues, i.e. the solution does not grow in the 2-norm.

The techniques developed by [89] and [96] were further analyzed and generalized by [97], with exponentially improved bounds on error for ill-conditioned diagonalizable linear ODE systems, non-diagonalizable systems of ODEs, and also provided an improved version of the algorithm provided by [45] for Carleman linearized nonlinear PDEs. [39] provide a theory of the overheads in the complexity of solving homogeneous ODEs using quantum computers compared to simulating quantum dynamics by identifying sources of "non-quantumness" in the problem and extend their results to inhomogeneous systems. They demonstrate that increased overheads arise from non-unique real parts of the eigenvalues of the ODE system, which can incur an exponential overhead in the worst case, and the non-orthogonality of the eigenvectors of the (Hermitian) ODE system measured as $\mu(A) = \left\| AA^\dagger - A^\dagger A \right\|^{1/2}$, which can incur an overhead linear in $\mu(A)$.

Quantum algorithms for inhomogeneous ODEs construct a larger linear system to encode all the time steps (or coefficients of the basis functions for the spectral approach) of the ODE using an additional register known as the Feynman-Kitaev clock and solve the resulting linear system. After solving the linear system, the desired iterate is extracted by measuring the Feynman-Kitaev register to obtain the final iterate with a non-zero probability of success. To boost the probability of getting the desired iterate, creating copies of the final state along with using amplitude amplification is a standard method. As an example, the method of [89] constructs a linear system of the form

$$
\begin{pmatrix}
I & & & & & & & & & & & & \\
-Ah & I & & & & & & & & & & & \\
& -\frac{Ah}{2} & I & & & & & & & & & & \\
& & \ddots & & & & & & & & & & \\
& & & -\frac{Ah}{k} & I & & & & & & & & \\
-I & -I & \cdots & -I & -I & I & & & & & & & \\
& & & & & -Ah & I & & & & & & \\
& & & & & & \ddots & & & & & & \\
& & & & & & & -\frac{Ah}{k} & I & & & & \\
& & & & & -I & \cdots & -I & -I & I & & & \\
& & & & & & & & & -I & I & & \\
& & & & & & & & & & \ddots & & \\
& & & & & & & & & & & -I & I
\end{pmatrix}
\begin{pmatrix}
\boldsymbol{x}_0 \\
Ah\boldsymbol{x}_0 \\
(Ah)^2/2\boldsymbol{x}_0 \\
\vdots \\
(Ah)^k/k!\,\boldsymbol{x}_0 \\
\boldsymbol{x}_1 \\
Ah\boldsymbol{x}_1 \\
\vdots \\
(Ah)^k/k!\,\boldsymbol{x}_{m-1} \\
\boldsymbol{x}_m \\
\boldsymbol{x}_m \\
\vdots \\
\boldsymbol{x}_m
\end{pmatrix}
=
\begin{pmatrix}
\boldsymbol{x}_0 \\
h\boldsymbol{b} \\
0 \\
\vdots \\
0 \\
0 \\
Ah\boldsymbol{x}_1 \\
\vdots \\
0 \\
0 \\
0 \\
\vdots \\
0
\end{pmatrix}
\tag{65}
$$

An algorithm with a complete circuit description and experimental results for a system of inhomogeneous ODEs of size $4 \times 4$ was given by [98] for unitary $\boldsymbol{A}$.

### 5.3.3.    Nonlinear Systems

Various algorithms and approaches have been presented for non-linear ODE problems. The first quantum algorithm for nonlinear differential equations (also the first quantum algorithm for any differential equation) was proposed by Leyton [99]. Leyton's algorithm proposed solving a nonlinear system of ordinary differential equations using the Euler method and consumes multiple copies of the initial condition, leading to an exponentially increasing cost in time. The algorithm requires the preparation of multiple copies of a quantum state to effect nonlinear transformations of amplitude encoded states. Using the block-encoding technique, nonlinear transformations of the singular values of a matrix $\boldsymbol{A}$ can be performed. However, nonlinear transformations of the amplitudes of a state as the input to a nonlinear function are not possible using this technique. [45] has provided an algorithm for dissipative nonlinear differential equations, specifically the $n$-dimensional quadratic ODE initial value problem, using the Carleman linearization technique. [100] use homotopic perturbation methods for exponentially improved precision for the homogeneous version of the algorithm in [45] with an orthogonal linear term. However, [97] points out that the problem considered by [100] has exponentially decaying solutions and therefore suffers from the post-selection problem. [101] explores the transformation of phase space to an equivalent Schrodinger equation using the Koopman-von Neumann formulation. [102] has provided a method for time-dependent problems using the Dyson series. As mentioned, the spectral approach of [96] can be used for time-dependent $\boldsymbol{A}(t)$. [103] has provided an algorithm for nonlinear differential equations using forward Euler time discretization for short time-intervals and also requires multiple copies of the initial condition, although in this case, the increase in cost is quadratic rather than exponential compared to [99]. A time-marching strategy for nonlinear ODEs has been proposed by [39], [95]. All quantum algorithms for nonlinear ODEs require the system to satisfy the condition of being dissipative.

### 5.4.    Quantum Algorithms for Partial Differential Equations

The solution of partial differential equations (PDEs) is of paramount importance due to their capability to simulate phenomena in many real-world problems. Partial differential equations of practical interest can typically be classified as elliptic, parabolic, and hyperbolic systems. PDEs are solved numerically on classical computers, with the finite element method, finite difference method, finite volume method, and spectral methods being popular techniques. These methods arrive at a discretized system of equations, which is then classically solved using either direct or iterative methods.

A quantum computer can be used to speed up the solution of a linear system of equations arising from a PDE discretization using QLSAs. However, as the number of unknowns increases, the linear system can become increasingly ill-conditioned. As an example, the condition number of a linear system arising from a finite element discretization can scale as $O(N^2)$, eliminating any speedup obtained from a direct application of a QLSA compared to classical iterative methods. Alternatively, some PDEs can be transformed into either the Schrodinger equation or a system of ODEs. Since quantum algorithms for Hamiltonian simulation and ODEs can provide exponential speedups with respect to the number of unknowns, they can also be used to solve PDEs. However, existing quantum algorithms using this approach are limited to structured grids on rectangular domains.

Various quantum algorithms have been proposed to solve PDEs using finite element [84], [92], finite volume [104], finite difference [111], [112], [113], and spectral methods [107]. Proposed approaches include solving the linear system arising from a discretization of the PDE [84], using Hamiltonian simulation to kick back phases [89], [105], mapping the PDE to the Schrodinger equation and evolving in time using Hamiltonian simulation [106], [108], and quantum ODE algorithms to evolve spatially discretized evolutionary PDEs [39]. We provide here an overview of these approaches for solving PDEs.

[84] proposes using QLSAs to solve the linear system arising from a finite element discretization with preconditioning using a sparse approximate-inverse preconditioner for an exponential speedup. This approach was further investigated by [92], who points out that when the cost of reading out the properties of the solution is included, the speedup is polynomial, with the speedup increasing for higher-dimensional problems. However, quantum circuits or procedures to implement the preconditioner were not provided.

### 5.4.1.    Schrodingerization

An algorithm to solve the Poisson problem with Dirichlet boundary conditions and rectangular grids was proposed by [105] with a linear scaling in $d$ and polylog scaling in $1/\epsilon$ for circuit size and a full circuit description. The algorithm uses the finite difference method to discretize the Laplacian operator on a unit cube domain and uses Hamiltonian simulation of the discretized operator. Due to the geometry of the domain, the Laplacian operator in $d$ dimensions can be expressed as the Kronecker sum

$$A = L_h \otimes I \otimes ... \otimes I + I \otimes L_h \otimes ... \otimes I + \cdots + I \otimes I \otimes ... \otimes L_h \tag{66}$$

where $L_h$ is the discretized Laplacian operator with grid spacing $h$. Using the exponentiation identity for Kronecker sums, the following Hamiltonian simulation is performed

$$e^{iAt} = e^{iL_h t} \otimes I \otimes ... \otimes I + I \otimes e^{iL_h t} \otimes ... \otimes I + \cdots + I \otimes I \otimes ... \otimes e^{iL_h t}. \tag{67}$$

Similar to the HHL algorithm, controlled versions of the Hamiltonian simulation are used to kick back the phase, and the remainder of the algorithm proceeds in the same manner. The algorithm of [105] was implemented with modifications for circuit optimization by [109] on a quantum simulator. [107] points out that while the circuit depth scales favorably, the probability of success is $O(\text{poly}(1/\epsilon))$ and finite-difference discretization errors are not considered in the analysis. [107] approaches the same problem with spectral and adaptive finite difference methods under a "global strict diagonal dominance" requirement, a stricter condition than diagonal dominance, to achieve a complexity of $O(d^2 \text{ poly} \log(1/\epsilon))$ using the spectral method and $O\left(d^{\frac{13}{2}} \text{ poly} \log(d/\epsilon)\right)$ using the adaptive finite difference grids where $d$ is the number of dimensions. However, the Kronecker product structure cannot be exploited for unstructured meshes on general domains encountered in problems of practical interest.

An algorithm for simulating the wave equation using the finite difference method was presented in [106] for Dirichlet and Neumann boundary conditions and with an approach based on a factorization of the Laplacian operator on a rectangular domain as $\boldsymbol{L}_h = \boldsymbol{B}\boldsymbol{B}^\dagger$ to transform the problem (2ⁿᵈ-order time derivative) into the Schrodinger equation (1ˢᵗ-order time derivative) solved using Hamiltonian simulation as

$$\frac{d}{dt}|\psi(t)\rangle = -\frac{i}{h}\begin{pmatrix} \boldsymbol{0} & \boldsymbol{B} \\ \boldsymbol{B}^\dagger & \boldsymbol{0} \end{pmatrix}|\psi(t)\rangle \tag{68}$$

by noting that

$$\frac{d^2}{dt^2}|\psi(t)\rangle = -\frac{1}{h^2}\begin{pmatrix} \boldsymbol{B}\boldsymbol{B}^\dagger & \boldsymbol{0} \\ \boldsymbol{0} & \boldsymbol{B}^\dagger\boldsymbol{B} \end{pmatrix}|\psi(t)\rangle = -\frac{1}{h^2}\begin{pmatrix} \boldsymbol{L}_h & \boldsymbol{0} \\ \boldsymbol{0} & \boldsymbol{L}_h{}^\dagger \end{pmatrix}|\psi(t)\rangle \tag{69}$$

where $|\psi(t)\rangle = \begin{pmatrix} \phi_V \\ \phi_E \end{pmatrix}$. $\phi_V$ encodes the solution to the wave equation and $\phi_E$ are intermediate variables. This algorithm was subsequently implemented by [110], albeit using the sub-optimal Trotter-Suzuki method for Hamiltonian simulation.

[108] provide a quantum algorithm for plasma physics problems. A linearized version of the Vlasov equation is derived, which is then evolved in time using Hamiltonian simulation algorithms similar to [106], and a complete circuit description with simulation results is provided with errors scaling as $O(\text{poly}(1/\epsilon))$. [111] uses a similar approach to map the cold plasma wave model to a Hamiltonian simulation problem but proposes using quantum signal processing for the Hamiltonian simulation step.

However, mapping a PDE to the Schrodinger equation has only been studied for these particular PDEs and may not be applicable to other PDEs.

### 5.4.2. Quantum ODE Solver Approaches

All algorithms for evolutionary PDEs mentioned above scale linearly in time, i.e., $O(t)$. [39] discretize hyperbolic and parabolic PDEs in space over a rectangular domain to obtain a system of ordinary differential equations. Quantum ODE solvers are then used to evolve the system in time instead of Hamiltonian simulation. They consider two scenarios to "fast-forward" (improve the time-complexity of) the simulation: semi-definite ODE systems with square-root access (similar to the decomposition in [106]) and diagonalizable systems (using a Fourier transform) to get $O(\sqrt{t})$ and $O(\log t)$ scaling in time. The lifting transformation proposed by [106] to transform a 2$^{\text{nd}}$-order ODE system $\frac{d^2}{dt^2}\boldsymbol{u}(t) = (\boldsymbol{A} + c\boldsymbol{I})\boldsymbol{u}(t) + \boldsymbol{b}(t)$ to a 1$^{\text{st}}$-order ODE system is extended for non-homogeneous systems as

$$\frac{d}{dt}\begin{pmatrix}\boldsymbol{u}(t)\\ \widetilde{\boldsymbol{v}}(t)\end{pmatrix} = \begin{pmatrix}\boldsymbol{0} & \boldsymbol{I}\\ (\boldsymbol{A}+c\boldsymbol{I}) & \boldsymbol{0}\end{pmatrix}\begin{pmatrix}\boldsymbol{u}(t)\\ \widetilde{\boldsymbol{v}}(t)\end{pmatrix} + \begin{pmatrix}\boldsymbol{0}\\ \boldsymbol{b}(t)\end{pmatrix}. \tag{70}$$

Analytical results for the transport equation, heat equation, advection-diffusion equation, wave equation, Klein-Gordon equation, Airy equation, and the Euler beam equation are provided. However, the fast-forwarding results do not apply to unstructured meshes on general domains.

Several algorithms have been proposed specifically for numerical solutions of the Navier-Stokes equations. Quantum lattice-gas models have been proposed by [118], [119], [120], [121], with numerical results presented in [116]. However, these methods are for Type-II quantum computers [117] whose architecture differs from the universal gate-based quantum computer architecture.

### 5.4.3. Quantum algorithms for the Navier-Stokes Equation

A quantum Navier-Stokes algorithm based on the quantum amplitude estimation algorithm was proposed by [118], which was later generalized to a quantum nonlinear PDE algorithm in [119]. However, the precision of the algorithms scales as $\Omega(\epsilon^{\frac{1}{q+1}})$, where $q = r + \rho$ is the smoothness parameter of the solution, with $r$ being the highest derivate retained in a Taylor series expansion and $0 \leq \rho \leq 1$, leading to a $O(\text{poly}(1/\epsilon))$ scaling. [120] improve the accuracy of the algorithm by an order of magnitude by using Chebyshev points but retain the $O(\text{poly}(1/\epsilon))$ scaling and provide numerical results using quantum simulators.

Lattice Boltzmann techniques also have the potential for quantum speedups of computational fluid dynamics. [121] has provided an approach with fully detailed circuits and simulator results for the solution of the advection-diffusion equation using the lattice Boltzmann method to achieve a scaling of $O(\log_2(\alpha D))$ where $\alpha$ is the total number of distribution functions and $D$ is the number of distribution functions for each site, albeit with a fixed choice of the relaxation time $\omega = 1$ and a limited choice of D1Q2 and D2Q5 models in 1D and 2D respectively. The approach requires classical computation for re-normalization of the post-selected state after each time step. We note, however, that [122] pointed out earlier that the streaming and collision operations are amenable to quantum computation and provide a quantum mapping of transport equations in fluid flows using analogies between the Dirac and Lattice Boltzmann equations to define an algorithm that measures an ancillary qubit at each time step, with a non-zero probability of success.

[123] provides an overview of techniques for mapping nonlinear systems to infinite-dimensional linear systems using Carleman embedding and truncating to finite dimensional systems for solution on quantum computers but does not provide an algorithm. [101] provides an overview of the Koopman-von Neumann method for a potential quadratic speedup.

## 6. Variational Algorithms and Quantum Machine Learning

With the exception of the Hamiltonian simulation of sparse systems using low-order Trotter-Suzuki methods and short evolution times for a small number of qubits, none of the algorithms discussed in the previous sections can be run on NISQ hardware due to its sensitivity to noise, imperfections in hardware fabrication, and low physical qubit counts preventing implementation of quantum error correction. In the NISQ era, there is great demand for algorithms that can provide quantum advantage using noisy hardware. Variational algorithms have recently gained attention as a hybrid classical-quantum approach for algorithms that can operate on NISQ devices. Several variational algorithms have been proposed with a wide range of applications, e.g., the Quantum Adiabatic Optimization Algorithm (QAOA) [124] for combinatorial problems, Variational Quantum Linear Solver (VQLS) [125] for linear systems, and the Variational Quantum Eigensolver [126] for preparing ground-states in quantum chemistry problems. The cost of a classical-quantum training loop prevents variational quantum algorithms from achieving a quantum advantage. Furthermore, the optimization or training of variational algorithms is specific to a problem and the quantum hardware and is not known to generalize. As an example, [29] has studied the efficacy of the VQLS in solving a 1D Poisson's problem, with the conclusion that the trained VQLS algorithm does not generalize to finer discretizations.

Variational quantum algorithms are remarkably similar to classical machine learning models, with the distinctions being the use of a quantum ansatz instead of a classical ansatz and the methods used for minimizing the cost function.

## 6.1.    Variational Quantum Algorithms

The ingredients of a variational quantum algorithm are a cost function $C(\boldsymbol{\theta})$, a parametrized quantum ansatz $U(\boldsymbol{\theta})$, and an optimization method (gradient-based or gradient-free) [127]. The cost function encodes the sought solution to a problem as:

$$\boldsymbol{\theta}^* = \arg\min_{\boldsymbol{\theta}} C(\boldsymbol{\theta}) \tag{71}$$

where $\boldsymbol{\theta}$ is a set of trainable parameters of an ansatz $U(\boldsymbol{\theta})$, a trainable quantum circuit parametrized by $\boldsymbol{\theta}$. The parameters $\boldsymbol{\theta}$ can be continuous or discrete. As an example, a continuous parameter could be the rotation angle for a $Y$ rotation gate, and a discrete parameter could be whether to apply a quantum gate in a circuit or not. The cost function can be expressed as

$$C(\boldsymbol{\theta}) = f\{\{\rho_k\}, \{O_k\}, U(\boldsymbol{\theta})\} \tag{72}$$

where $f$ is some function, $\{\rho_k\}$ are input states corresponding to training data, and $\{O_k\}$ are observables, or measurements of the quantum circuit. The evaluation of $C(\boldsymbol{\theta})$ is performed using quantum computers (possibly with some classical post-processing), and the optimization of $\boldsymbol{\theta}$ is performed using classical computers, leading to a hybrid algorithm. For NISQ hardware, the quantum circuit for evaluation of $C(\boldsymbol{\theta})$ should be within the constraints of available hardware. This entails having small numbers of qubits and circuit depths.

The choice of the ansatz determines the parameters $\boldsymbol{\theta}$. In general, a quantum ansatz can be expressed as

$$U(\boldsymbol{\theta}) = \prod_{l=1}^{L} U_l(\theta_l). \tag{73}$$

## 6.2.    Ansatz parameter optimization

Various ansatze have been proposed for a variety of problems with the unifying goal of being efficient and trainable on NISQ hardware.

The cost function may be optimized using gradient-based or gradient-free approaches. Gradient-based approaches compute the derivatives $\frac{\partial C(\theta_l)}{\partial \theta_l}$.

The derivative may be approximated using finite differencing. However, "parameter shift" rules [134], [135], [136], [137] allow exact differentiation of parametrized quantum circuits (for continuous parameters) analogous to automatic differentiation for classical computing. The key idea behind parameter shift rules is that to compute $\frac{\partial C(\theta_l)}{\partial \theta_l}$, the same quantum circuit is used with a "parameter shift" $s \in \mathbb{R}$ applied to the parameter $\theta_l$ with a multiplier $c \in \mathbb{R}$:

$$\frac{\partial C(\boldsymbol{\theta})}{\partial \theta_l} = c\left[C(\boldsymbol{\theta}^{(l)}) - C(\boldsymbol{\theta}^{(l)})\right] \tag{74}$$

where

$$\theta_k^{(l)} = \begin{cases} \theta_k, & \forall\, k \neq l \\ \theta_k + s & k = l. \end{cases} \tag{75}$$

Although the parameter shift rule resembles finite differences, it is exact. Higher-order derivatives may be computed by nesting the parameter shift rule. Parameter shift rules can be computed for any arbitrary circuit with continuous parameters. Furthermore, a quantum ansatz can be combined with classical ansatz using a hybrid of parameter shift rules for the quantum circuit and automatic differentiation for classical circuits [127], [132].

However, optimization of quantum circuits can suffer from vanishing gradients [132], [133]. In these cases, classical derivative-free optimizers like Nelder-Mead [134], [135] can be used as an alternative if the number of parameters is modest.

### 6.3.    Notable variational algorithms

The most notable variational quantum algorithm is the variational quantum eigensolver (VQE), which was introduced by [136] to prepare the groundstate of the helium hydride ion. The VQE has since been used to compute and prepare the ground and excited states and energies of more complex molecules [126], [137], [138].

Variational quantum linear solvers (VQLS) [125] have also gained attention recently. [29], [30] use the VQLS to solve the heat equation in 1D, [139] provides a solution in 2D, and [140] has demonstrated a solution of potential and Stokes flow in 2D. [139], [140] demonstrate logarithmic scaling in $1/\epsilon$ and $N$. However, we note that they use the Pauli basis to represent their matrix, which can have $O(N)$ terms for a 1D discrete Laplacian. [141] provides an efficient tensor product decomposition with $O(\log(N))$ terms.

The quantum approximate optimization algorithm (QAOA) attempts to find the ground-state of the Ising Hamiltonian. Since quantum annealing primarily uses the Ising Hamiltonian, algorithms developed for quantum annealers can be ported to gate-based quantum computers. [83] implement the time optimal QLSA of [81] using the QAOA using an optimally tuned scheduling function and demonstrate near-optimal performance using numerical experiments.

[142] has developed a variational fast-forwarding technique to fast-forward Hamiltonian simulation by using a Trotter circuit to train a fast-forwardable variational ansatz to drastically reduce the simulation time, therefore allowing longer Hamiltonian simulations to be performed on NISQ hardware.

Much like classical machine learning, variational algorithms have been explored for a wide variety of problems. [143], [144] explore variational algorithms for stochastic PDEs in finance. A thorough review conducted by JPMorgan Chase for financial applications can be found in [145]. [146] has proposed a variational algorithm for cosmological simulations.

Variational algorithms have also been proposed for nonlinear PDEs. [147] uses multiple copies of variational quantum states to treat nonlinearities and numerically demonstrates that variational ansatz can be exponentially more efficient than matrix product states and presents experimental results as a proof-of-concept. [148] presents a Chebyshev feature map for nonlinear PDEs and provides simulation results for the Navier Stokes to compute flow and fluid properties.

## 7. Roadmap & Future Directions

As the research area of quantum computing is in its infancy, there remain many unexplored avenues and problems, and quantum scientific computing in the past decade has made tremendous progress. However, error correction is a major challenge to develop logical qubits with long coherence times. Although quantum algorithms can demonstrate great speedups for various problems, practical implementation of quantum algorithms requires a great deal of research effort. There is great potential for quantum speedups for scientific computing, and many problems that need to be tackled. The design of circuits of oracles for practically relevant problems that are not only efficient in the number of gates and ancilla qubits but also in the sub-normalization constants is also crucial to implementing quantum algorithms for scientific computing beyond the NISQ era. Quantum algorithms with full circuit descriptions are rare; implementation of algorithms and numerical results can give important insights into the comparative feasibility of various algorithms and bring to attention potential bottlenecks. Furthermore, circuit implementations of problems and matrices can usher in guidelines or "rules-of-thumb" for circuit design and, in some cases, provide a modular solution.

Efficient circuits for qubitizations of unstructured mesh finite element problems have not been developed. Although the proposition of sparse access models on gate-based quantum computers is that locally and efficiently computable entries of a matrix can be accessed efficiently, i.e., computed directly on a quantum computer, the computational infrastructure for these access models is nonexistent. Furthermore, as fault-tolerant quantum computers are developed, the earliest generations are unlikely to be capable of handling the entire computational pipeline starting from a geometric model and outputting properties of the finite element solution. For NISQ hardware, quantum circuits specific to a problem may need to be generated, but this can be costly. A scheme to use a quantum circuit that qubitizes a coarse unstructured mesh to generate a qubitization of a fine mesh would be valuable since it could be used recursively to perform mesh refinement without forming a quantum circuit for a fine mesh separately.

Current algorithms for quantum computing lie on opposite ends of the fault-tolerance spectrum. Variational algorithms can produce coherent results on NISQ hardware but do not demonstrate generalization or scalability to larger problems, while algorithms demonstrating exponential speedups require long hardware coherence times and low error rates. New algorithms need to be developed that can operate in the intermediate regime. Trotter-based algorithms require fewer qubits and shorter circuits and are, therefore, expected to be the first algorithms to provide quantum speedups [7]. Error extrapolation and mitigation techniques are also an active area of research to improve performance with NISQ hardware [149].

However, in the NISQ era, the search for variational algorithms for quantum advantage (Daley et al., 2022) or quantum utility (Kim et al., 2023) is likely to dominate the field. However, it is difficult to prove a speedup compared to classical algorithms. Hardware development for quantum computers is likely to be driven by early applications which includes scientific computing. One may expect QPUs specialized for Trotter methods or similar NISQ or utility era computing tasks.

## Acknowledgements

The authors would like to acknowledge the support of this work by NIBIB R01EB0005807, R01EB25241, R01EB033674 and R01EB032820 grants.

# References

[1]  M. A. Nielsen and I. L. Chuang, *Quantum Computation and Quantum Information: 10th Anniversary Edition*, 1st ed. Cambridge University Press, 2012. doi: 10.1017/CBO9780511976667.

[2]  R. P. Feynman, "Quantum mechanical computers," *Found Phys*, vol. 16, no. 6, pp. 507–531, Jun. 1986, doi: 10.1007/BF01886518.

[3]  P. W. Shor, "Algorithms for quantum computation: discrete logarithms and factoring," in *Proceedings 35th Annual Symposium on Foundations of Computer Science*, Santa Fe, NM, USA: IEEE Comput. Soc. Press, 1994, pp. 124–134. doi: 10.1109/SFCS.1994.365700.

[4]  A. M. Childs, R. Kothari, and R. D. Somma, "Quantum algorithm for systems of linear equations with exponentially improved dependence on precision," *SIAM J. Comput.*, vol. 46, no. 6, pp. 1920–1950, Jan. 2017, doi: 10.1137/16M1087072.

[5]  A. W. Harrow, A. Hassidim, and S. Lloyd, "Quantum algorithm for linear systems of equations," *Phys. Rev. Lett.*, vol. 103, no. 15, p. 150502, Oct. 2009, doi: 10.1103/PhysRevLett.103.150502.

[6]  D. Aharonov, W. van Dam, J. Kempe, Z. Landau, S. Lloyd, and O. Regev, "Adiabatic quantum computation is equivalent to standard quantum computation," *SIAM Review*, vol. 50, no. 4, pp. 755–787, 2008.

[7]  A. M. Childs, D. Maslov, Y. Nam, N. J. Ross, and Y. Su, "Toward the first quantum simulation with quantum speedup," *Proc. Natl. Acad. Sci. U.S.A.*, vol. 115, no. 38, pp. 9456–9461, Sep. 2018, doi: 10.1073/pnas.1801723115.

[8]  Google Quantum AI *et al.*, "Exponential suppression of bit or phase errors with cyclic error correction," *Nature*, vol. 595, no. 7867, pp. 383–387, Jul. 2021, doi: 10.1038/s41586-021-03588-y.

[9]  Google Quantum AI *et al.*, "Suppressing quantum errors by scaling a surface code logical qubit," *Nature*, vol. 614, no. 7949, pp. 676–681, Feb. 2023, doi: 10.1038/s41586-022-05434-1.

[10]  A. Gonzales, R. Shaydulin, Z. H. Saleem, and M. Suchara, "Quantum error mitigation by Pauli check sandwiching," *Sci Rep*, vol. 13, no. 1, p. 2122, Feb. 2023, doi: 10.1038/s41598-023-28109-x.

[11]  R. Majumdar, P. Rivero, F. Metz, A. Hasan, and D. S. Wang, "Best Practices for Quantum Error Mitigation with Digital Zero-Noise Extrapolation," in *2023 IEEE International Conference on Quantum Computing and Engineering (QCE)*, Bellevue, WA, USA: IEEE, Sep. 2023, pp. 881–887. doi: 10.1109/QCE57702.2023.00102.

[12]  J. Qi, X. Xu, D. Poletti, and H. K. Ng, "Efficacy of noisy dynamical decoupling," *Phys. Rev. A*, vol. 107, no. 3, p. 032615, Mar. 2023, doi: 10.1103/PhysRevA.107.032615.

[13]  J. Robledo-Moreno *et al.*, "Chemistry Beyond Exact Solutions on a Quantum-Centric Supercomputer," 2024, *arXiv*. doi: 10.48550/ARXIV.2405.05068.

[14]  M. Born, "Zur quantenmechanik der stoßvorgänge," *Z. Physik*, vol. 37, no. 12, pp. 863–867, Dec. 1926, doi: 10.1007/BF01397477.

[15]  N. P. Landsman, "Born rule and its interpretation," in *Compendium of Quantum Physics*, D. Greenberger, K. Hentschel, and F. Weinert, Eds., Berlin, Heidelberg: Springer Berlin

Heidelberg, 2009, pp. 64–70. doi: 10.1007/978-3-540-70626-7_20.

[16]  A. M. Childs, "Universal computation by quantum walk," *Phys. Rev. Lett.*, vol. 102, no. 18, p. 180501, May 2009, doi: 10.1103/PhysRevLett.102.180501.

[17]  M. Weigold, J. Barzen, F. Leymann, and M. Salm, "Encoding patterns for quantum algorithms," *IET Quantum Communication*, vol. 2, no. 4, pp. 141–152, Dec. 2021, doi: 10.1049/qtc2.12032.

[18]  L. Lin, "Lecture notes on quantum algorithms for scientific computation," 2022, *arXiv*. doi: 10.48550/ARXIV.2201.08309.

[19]  *IEEE Standard for Floating-Point Arithmetic*, Jul. 22, 2019. doi: 10.1109/IEEESTD.2019.8766229.

[20]  A. Y. Kitaev, "Quantum computations: algorithms and error correction," *Russ. Math. Surv.*, vol. 52, no. 6, pp. 1191–1249, Dec. 1997, doi: 10.1070/RM1997v052n06ABEH002155.

[21]  C. M. Dawson and M. A. Nielsen, "The Solovay-Kitaev algorithm," *Quantum Info. Comput.*, vol. 6, no. 1, pp. 81–95, Jan. 2006.

[22]  J. B. Altepeter, D. F. V. James, and P. G. Kwiat, "Qubit quantum state tomography," in *Quantum State Estimation*, vol. 649, M. Paris and J. Řeháček, Eds., in Lecture Notes in Physics, vol. 649. , Berlin, Heidelberg: Springer Berlin Heidelberg, 2004, pp. 113–145. doi: 10.1007/978-3-540-44481-7_4.

[23]  Qiskit contributors, *Qiskit: An Open-source Framework for Quantum Computing*. (Jul. 27, 2023). doi: 10.5281/ZENODO.2573505.

[24]  V. Bergholm *et al.*, "PennyLane: Automatic differentiation of hybrid quantum-classical computations," 2018, *arXiv*. doi: 10.48550/ARXIV.1811.04968.

[25]  S. Chakraborty, A. Gilyén, and S. Jeffery, "The power of block-encoded matrix powers: improved regression techniques via faster Hamiltonian simulation," in *Leibniz International Proceedings in Informatics*, Schloss Dagstuhl - Leibniz-Zentrum für Informatik, 2019, p. 33:1-33:14. doi: 10.4230/LIPICS.ICALP.2019.33.

[26]  J. M. Martyn, Z. M. Rossi, A. K. Tan, and I. L. Chuang, "Grand unification of quantum algorithms," *PRX Quantum*, vol. 2, no. 4, p. 040203, Dec. 2021, doi: 10.1103/PRXQuantum.2.040203.

[27]  D. Camps, L. Lin, R. Van Beeumen, and C. Yang, "Explicit quantum circuits for block encodings of certain sparse matrices," 2022, *arXiv*. doi: 10.48550/ARXIV.2203.10236.

[28]  A. C. Vazquez, "Quantum Algorithma for solving tri-diagonal linear systems of equations," ETH Zurich, 2018. [Online]. Available: http://www.sam.math.ethz.ch/~hiptmair/StudentProjects/CarreraVazquez.Almudena/MScThesis.pdf

[29]  E. Cappanera, "Variational quantum linear solver for finite element problems: a Poisson equation test case," TU Delft, 2021. [Online]. Available: http://resolver.tudelft.nl/uuid:deba389d-f30f-406c-ad7b-babb1b298d87

[30]  C. J. Trahan, M. Loveland, N. Davis, and E. Ellison, "A variational quantum linear solver application to discrete finite-element methods," *Entropy*, vol. 25, no. 4, p. 580, Mar. 2023, doi: 10.3390/e25040580.

[31]  S. Chakraborty, A. Morolia, and A. Peduri, "Quantum regularized least squares," *Quantum*, vol. 7, p. 988, Apr. 2023, doi: 10.22331/q-2023-04-27-988.

[32]  A. Gilyén, Y. Su, G. H. Low, and N. Wiebe, "Quantum singular value transformation and

beyond: exponential improvements for quantum matrix arithmetics," in *Proceedings of the 51st Annual ACM SIGACT Symposium on Theory of Computing*, Phoenix AZ USA: ACM, Jun. 2019, pp. 193–204. doi: 10.1145/3313276.3316366.

[33] D. Camps and R. Van Beeumen, "Approximate quantum circuit synthesis using block encodings," *Phys. Rev. A*, vol. 102, no. 5, p. 052411, Nov. 2020, doi: 10.1103/PhysRevA.102.052411.

[34] D. W. Berry, A. M. Childs, R. Cleve, R. Kothari, and R. D. Somma, "Exponential improvement in precision for simulating sparse Hamiltonians," in *Proceedings of the Forty-Sixth Annual ACM Symposium on Theory of Computing*, in STOC '14. New York, NY, USA: Association for Computing Machinery, 2014, pp. 283–292. doi: 10.1145/2591796.2591854.

[35] W. K. Wootters and W. H. Zurek, "A single quantum cannot be cloned," *Nature*, vol. 299, no. 5886, pp. 802–803, Oct. 1982, doi: 10.1038/299802a0.

[36] V. Bužek and M. Hillery, "Quantum copying: Beyond the no-cloning theorem," *Phys. Rev. A*, vol. 54, no. 3, pp. 1844–1852, Sep. 1996, doi: 10.1103/PhysRevA.54.1844.

[37] V. Bužek and M. Hillery, "Universal optimal cloning of arbitrary quantum states: From qubits to quantum registers," *Phys. Rev. Lett.*, vol. 81, no. 22, pp. 5003–5006, Nov. 1998, doi: 10.1103/PhysRevLett.81.5003.

[38] V. Buzek and M. Hillery, "Quantum cloning," *Phys. World*, vol. 14, no. 11, pp. 25–30, Nov. 2001, doi: 10.1088/2058-7058/14/11/28.

[39] D. An, J.-P. Liu, D. Wang, and Q. Zhao, "A theory of quantum differential equation solvers: limitations and fast-forwarding," 2022, *arXiv*. doi: 10.48550/ARXIV.2211.05246.

[40] G. Brassard, P. Høyer, M. Mosca, and A. Tapp, "Quantum amplitude amplification and estimation," in *Contemporary Mathematics*, vol. 305, S. J. Lomonaco and H. E. Brandt, Eds., Providence, Rhode Island: American Mathematical Society, 2002, pp. 53–74. doi: 10.1090/conm/305/05215.

[41] M. Cramer *et al.*, "Efficient quantum state tomography," *Nat Commun*, vol. 1, no. 1, p. 149, Dec. 2010, doi: 10.1038/ncomms1147.

[42] K. Phalak, A. Chatterjee, and S. Ghosh, "Quantum random access memory For dummies," 2023, *arXiv*. doi: 10.48550/ARXIV.2305.01178.

[43] D. W. Berry, G. Ahokas, R. Cleve, and B. C. Sanders, "Efficient quantum algorithms for simulating sparse Hamiltonians," *Commun. Math. Phys.*, vol. 270, no. 2, pp. 359–371, Mar. 2007, doi: 10.1007/s00220-006-0150-x.

[44] E. Farhi, J. Goldstone, S. Gutmann, and M. Sipser, "Limit on the speed of quantum computation in determining parity," *Phys. Rev. Lett.*, vol. 81, no. 24, pp. 5442–5444, Dec. 1998, doi: 10.1103/PhysRevLett.81.5442.

[45] J.-P. Liu, H. Ø. Kolden, H. K. Krovi, N. F. Loureiro, K. Trivisa, and A. M. Childs, "Efficient quantum algorithm for dissipative nonlinear differential equations," *Proc. Natl. Acad. Sci. U.S.A.*, vol. 118, no. 35, p. e2026805118, Aug. 2021, doi: 10.1073/pnas.2026805118.

[46] M. A. Schalkers and M. Möller, "On the importance of data encoding in quantum Boltzmann methods," Feb. 10, 2023, *arXiv*: arXiv:2302.05305. Accessed: Nov. 07, 2023. [Online]. Available: http://arxiv.org/abs/2302.05305

[47] Y. Liu, S. J. Srinivasan, D. Hover, S. Zhu, R. McDermott, and A. A. Houck, "High fidelity readout of a transmon qubit using a superconducting low-inductance undulatory galvanometer microwave amplifier," *New J. Phys.*, vol. 16, no. 11, p. 113008, Nov. 2014,

doi: 10.1088/1367-2630/16/11/113008.

[48] P. Kok, W. J. Munro, K. Nemoto, T. C. Ralph, J. P. Dowling, and G. J. Milburn, "Linear optical quantum computing with photonic qubits," *Rev. Mod. Phys.*, vol. 79, no. 1, pp. 135–174, Jan. 2007, doi: 10.1103/RevModPhys.79.135.

[49] A. Cross *et al.*, "OpenQASM 3: A broader and deeper quantum assembly language," *ACM Transactions on Quantum Computing*, vol. 3, no. 3, pp. 1–50, Sep. 2022, doi: 10.1145/3505636.

[50] N. Khammassi, G. G. Guerreschi, I. Ashraf, J. W. Hogaboam, C. G. Almudever, and K. Bertels, "cQASM v1.0: Towards a common quantum assembly language," 2018, *arXiv*. doi: 10.48550/ARXIV.1805.09607.

[51] B. C. A. Morrison *et al.*, "Just another quantum assembly language (Jaqal)," in *2020 IEEE International Conference on Quantum Computing and Engineering (QCE)*, Denver, CO, USA: IEEE, Oct. 2020, pp. 402–408. doi: 10.1109/QCE49297.2020.00056.

[52] Cirq Developers, *Cirq*. (Jul. 18, 2023). Zenodo. doi: 10.5281/ZENODO.4062499.

[53] J. Hooyberghs, *Introducing Microsoft Quantum Computing for Developers: Using the Quantum Development Kit and Q#*. Berkeley, CA: Apress, 2022. doi: 10.1007/978-1-4842-7246-6.

[54] N. Killoran, J. Izaac, N. Quesada, V. Bergholm, M. Amy, and C. Weedbrook, "Strawberry Fields: A software platform for photonic quantum computing," *Quantum*, vol. 3, p. 129, Mar. 2019, doi: 10.22331/q-2019-03-11-129.

[55] P. J. Karalekas *et al.*, *PyQuil: Quantum programming in Python*. (Jan. 30, 2020). Zenodo. doi: 10.5281/ZENODO.3553165.

[56] J. R. McClean *et al.*, "OpenFermion: the electronic structure package for quantum computers," *Quantum Sci. Technol.*, vol. 5, no. 3, p. 034014, Jun. 2020, doi: 10.1088/2058-9565/ab8ebc.

[57] Q. Sun *et al.*, "Recent developments in the PySCF program package," *The Journal of Chemical Physics*, vol. 153, no. 2, p. 024109, Jul. 2020, doi: 10.1063/5.0006074.

[58] P. Jordan and E. Wigner, "Über das Paulische äquivalenzverbot," *Z. Physik*, vol. 47, no. 9–10, pp. 631–651, Sep. 1928, doi: 10.1007/BF01331938.

[59] S. B. Bravyi and A. Yu. Kitaev, "Fermionic quantum computation," *Annals of Physics*, vol. 298, no. 1, pp. 210–226, May 2002, doi: 10.1006/aphy.2002.6254.

[60] Y. Dong, X. Meng, K. B. Whaley, and L. Lin, "Efficient phase-factor evaluation in quantum signal processing," *Phys. Rev. A*, vol. 103, no. 4, p. 042419, Apr. 2021, doi: 10.1103/PhysRevA.103.042419.

[61] S. Günther, N. A. Petersson, and J. L. DuBois, "Quantum optimal control for pure-state preparation using one initial state," *AVS Quantum Sci.*, vol. 3, no. 4, p. 043801, Dec. 2021, doi: 10.1116/5.0060262.

[62] Y. Nam, N. J. Ross, Y. Su, A. M. Childs, and D. Maslov, "Automated optimization of large quantum circuits with continuous parameters," *npj Quantum Inf*, vol. 4, no. 1, p. 23, May 2018, doi: 10.1038/s41534-018-0072-4.

[63] T. Giurgica-Tiron, Y. Hindy, R. LaRose, A. Mari, and W. J. Zeng, "Digital zero noise extrapolation for quantum error mitigation," in *2020 IEEE International Conference on Quantum Computing and Engineering (QCE)*, Denver, CO, USA: IEEE, Oct. 2020, pp. 306–316. doi: 10.1109/QCE49297.2020.00045.

[64] C. Coppersmith, "An approximate Fourier transform useful in quantum factoring," IBM Research Division, RC 19642, Jun. 1994. [Online]. Available: https://doi.org/10.48550/arXiv.quant-ph/0201067

[65] C. J. O'Loan, "Iterative phase estimation," *J. Phys. A: Math. Theor.*, vol. 43, no. 1, p. 015301, Jan. 2010, doi: 10.1088/1751-8113/43/1/015301.

[66] A. N. Chowdhury, Y. Subasi, and R. D. Somma, "Improved implementation of reflection operators," 2018, *arXiv*. doi: 10.48550/ARXIV.1803.02466.

[67] A. Ambainis, "Variable time amplitude amplification and quantum algorithms for linear algebra problems," in *Leibniz International Proceedings in Informatics*, Schloss Dagstuhl - Leibniz-Zentrum für Informatik, 2012, pp. 636–647. doi: 10.4230/LIPICS.STACS.2012.636.

[68] L. K. Grover, "A fast quantum mechanical algorithm for database search," in *Proceedings of the twenty-eighth annual ACM symposium on Theory of computing - STOC '96*, Philadelphia, Pennsylvania, United States: ACM Press, 1996, pp. 212–219. doi: 10.1145/237814.237866.

[69] A. M. Childs and N. Wiebe, "Hamiltonian simulation using linear combinations of unitary operations," *QIC*, vol. 12, no. 11 & 12, Nov. 2012, doi: 10.26421/QIC12.11-12.

[70] G. H. Low and I. L. Chuang, "Hamiltonian simulation by qubitization," *Quantum*, vol. 3, p. 163, Jul. 2019, doi: 10.22331/q-2019-07-12-163.

[71] G. H. Low and I. L. Chuang, "Optimal Hamiltonian simulation by quantum signal processing," *Phys. Rev. Lett.*, vol. 118, no. 1, p. 010501, Jan. 2017, doi: 10.1103/PhysRevLett.118.010501.

[72] D. Motlagh and N. Wiebe, "Generalized quantum signal processing," 2023, *arXiv*. doi: 10.48550/ARXIV.2308.01501.

[73] C.-F. Chen, H.-Y. Huang, R. Kueng, and J. A. Tropp, "Concentration for random product formulas," *PRX Quantum*, vol. 2, no. 4, p. 040305, Oct. 2021, doi: 10.1103/PRXQuantum.2.040305.

[74] A. M. Childs, A. Ostrander, and Y. Su, "Faster quantum simulation by randomization," *Quantum*, vol. 3, p. 182, Sep. 2019, doi: 10.22331/q-2019-09-02-182.

[75] A. Smith, M. S. Kim, F. Pollmann, and J. Knolle, "Simulating quantum many-body dynamics on a current digital quantum computer," *npj Quantum Inf*, vol. 5, no. 1, p. 106, Nov. 2019, doi: 10.1038/s41534-019-0217-0.

[76] M. Suzuki, "Generalized Trotter's formula and systematic approximants of exponential operators and inner derivations with applications to many-body problems," *Commun.Math. Phys.*, vol. 51, no. 2, pp. 183–190, Jun. 1976, doi: 10.1007/BF01609348.

[77] J. D. Whitfield, J. Biamonte, and A. Aspuru-Guzik, "Simulation of electronic structure Hamiltonians using quantum computers," *Molecular Physics*, vol. 109, no. 5, pp. 735–750, Mar. 2011, doi: 10.1080/00268976.2011.552441.

[78] M. E. S. Morales, P. C. S. Costa, D. K. Burgarth, Y. R. Sanders, and D. W. Berry, "Greatly improved higher-order product formulae for quantum simulation," 2022, *arXiv*. doi: 10.48550/ARXIV.2210.15817.

[79] A. Schubert and C. B. Mendl, "Trotter error with commutator scaling for the Fermi-Hubbard model," 2023, *arXiv*. doi: 10.48550/ARXIV.2306.10603.

[80] D. W. Berry, A. M. Childs, R. Cleve, R. Kothari, and R. D. Somma, "Simulating

Hamiltonian dynamics with a yruncated Taylor series," *Phys. Rev. Lett.*, vol. 114, no. 9, p. 090502, Mar. 2015, doi: 10.1103/PhysRevLett.114.090502.

[81] P. C. S. Costa, D. An, Y. R. Sanders, Y. Su, R. Babbush, and D. W. Berry, "Optimal scaling quantum linear-systems solver via discrete adiabatic theorem," *PRX Quantum*, vol. 3, no. 4, p. 040303, Oct. 2022, doi: 10.1103/PRXQuantum.3.040303.

[82] D. Orsucci and V. Dunjko, "On solving classes of positive-definite quantum linear systems with quadratically improved runtime in the condition number," *Quantum*, vol. 5, p. 573, Nov. 2021, doi: 10.22331/q-2021-11-08-573.

[83] D. An and L. Lin, "Quantum linear system solver based on time-optimal adiabatic quantum computing and quantum approximate optimization algorithm," *ACM Transactions on Quantum Computing*, vol. 3, no. 2, pp. 1–28, Jun. 2022, doi: 10.1145/3498331.

[84] B. D. Clader, B. C. Jacobs, and C. R. Sprouse, "Preconditioned quantum linear system algorithm," *Phys. Rev. Lett.*, vol. 110, no. 25, p. 250504, Jun. 2013, doi: 10.1103/PhysRevLett.110.250504.

[85] J. Golden, D. O'Malley, and H. Viswanathan, "Quantum computing and preconditioners for hydrological linear systems," *Sci Rep*, vol. 12, no. 1, p. 22285, Dec. 2022, doi: 10.1038/s41598-022-25727-9.

[86] C. Shao and H. Xiang, "Quantum circulant preconditioner for a linear system of equations," *Phys. Rev. A*, vol. 98, no. 6, p. 062321, Dec. 2018, doi: 10.1103/PhysRevA.98.062321.

[87] Y. Tong, D. An, N. Wiebe, and L. Lin, "Fast inversion, preconditioned quantum linear system solvers, fast Green's-function computation, and fast evaluation of matrix functions," *Phys. Rev. A*, vol. 104, no. 3, p. 032422, Sep. 2021, doi: 10.1103/PhysRevA.104.032422.

[88] D. W. Berry, "High-order quantum algorithm for solving linear differential equations," *J. Phys. A: Math. Theor.*, vol. 47, no. 10, p. 105301, Mar. 2014, doi: 10.1088/1751-8113/47/10/105301.

[89] D. W. Berry, A. M. Childs, A. Ostrander, and G. Wang, "Quantum algorithm for linear differential equations with exponentially improved dependence on precision," *Commun. Math. Phys.*, vol. 356, no. 3, pp. 1057–1081, Dec. 2017, doi: 10.1007/s00220-017-3002-y.

[90] O. M. Raisuddin and S. De, "Quantum relaxation for linear systems in finite element analysis," 2023, *arXiv*. doi: 10.48550/ARXIV.2308.01377.

[91] O. M. Raisuddin and S. De, "Quantum Multigrid Algorithm for Finite Element Problems," 2024, *arXiv*. doi: 10.48550/ARXIV.2404.07466.

[92] A. Montanaro and S. Pallister, "Quantum algorithms and the finite element method," *Phys. Rev. A*, vol. 93, no. 3, p. 032324, Mar. 2016, doi: 10.1103/PhysRevA.93.032324.

[93] N. Linden, A. Montanaro, and C. Shao, "Quantum vs. classical algorithms for solving the heat equation," *Commun. Math. Phys.*, vol. 395, no. 2, pp. 601–641, Oct. 2022, doi: 10.1007/s00220-022-04442-6.

[94] S. Lloyd, "Universal quantum simulators," *Science*, vol. 273, no. 5278, pp. 1073–1078, Aug. 1996, doi: 10.1126/science.273.5278.1073.

[95] D. Fang, L. Lin, and Y. Tong, "Time-marching based quantum solvers for time-dependent linear differential equations," *Quantum*, vol. 7, p. 955, Mar. 2023, doi: 10.22331/q-2023-03-20-955.

[96] A. M. Childs and J.-P. Liu, "Quantum spectral methods for differential equations," *Commun. Math. Phys.*, vol. 375, no. 2, pp. 1427–1457, Apr. 2020, doi: 10.1007/s00220-020-03699-z.

[97] H. Krovi, "Improved quantum algorithms for linear and nonlinear differential equations," *Quantum*, vol. 7, p. 913, Feb. 2023, doi: 10.22331/q-2023-02-02-913.

[98] T. Xin *et al.*, "Quantum algorithm for solving linear differential equations: Theory and experiment," *Phys. Rev. A*, vol. 101, no. 3, p. 032307, Mar. 2020, doi: 10.1103/PhysRevA.101.032307.

[99] S. K. Leyton and T. J. Osborne, "A quantum algorithm to solve nonlinear differential equations," 2008, *arXiv*. doi: 10.48550/ARXIV.0812.4423.

[100] C. Xue, Y.-C. Wu, and G.-P. Guo, "Quantum homotopy perturbation method for nonlinear dissipative ordinary differential equations," *New J. Phys.*, vol. 23, no. 12, p. 123035, Dec. 2021, doi: 10.1088/1367-2630/ac3eff.

[101] I. Joseph, "Koopman–von Neumann approach to quantum simulation of nonlinear classical dynamics," *Phys. Rev. Research*, vol. 2, no. 4, p. 043102, Oct. 2020, doi: 10.1103/PhysRevResearch.2.043102.

[102] D. W. Berry and P. C. S. Costa, "Quantum algorithm for time-dependent differential equations using Dyson series," 2022, *arXiv*. doi: 10.48550/ARXIV.2212.03544.

[103] S. Lloyd *et al.*, "Quantum algorithm for nonlinear differential equations," 2020, *arXiv*. doi: 10.48550/ARXIV.2011.06571.

[104] F. Fillion-Gourdeau and E. Lorin, "Simple digital quantum algorithm for symmetric first-order linear hyperbolic systems," *Numer Algor*, vol. 82, no. 3, pp. 1009–1045, Nov. 2019, doi: 10.1007/s11075-018-0639-3.

[105] Y. Cao, A. Papageorgiou, I. Petras, J. Traub, and S. Kais, "Quantum algorithm and circuit design solving the Poisson equation," *New J. Phys.*, vol. 15, no. 1, p. 013021, Jan. 2013, doi: 10.1088/1367-2630/15/1/013021.

[106] P. C. S. Costa, S. Jordan, and A. Ostrander, "Quantum algorithm for simulating the wave equation," *Phys. Rev. A*, vol. 99, no. 1, p. 012323, Jan. 2019, doi: 10.1103/PhysRevA.99.012323.

[107] A. M. Childs, J.-P. Liu, and A. Ostrander, "High-precision quantum algorithms for partial differential equations," *Quantum*, vol. 5, p. 574, Nov. 2021, doi: 10.22331/q-2021-11-10-574.

[108] A. Engel, G. Smith, and S. E. Parker, "Quantum algorithm for the Vlasov equation," *Phys. Rev. A*, vol. 100, no. 6, p. 062315, Dec. 2019, doi: 10.1103/PhysRevA.100.062315.

[109] S. Wang, Z. Wang, W. Li, L. Fan, Z. Wei, and Y. Gu, "Quantum fast Poisson solver: the algorithm and complete and modular circuit design," *Quantum Inf Process*, vol. 19, no. 6, p. 170, Jun. 2020, doi: 10.1007/s11128-020-02669-7.

[110] A. Suau, G. Staffelbach, and H. Calandra, "Practical quantum computing: Solving the wave equation using a quantum approach," *ACM Transactions on Quantum Computing*, vol. 2, no. 1, pp. 1–35, Mar. 2021, doi: 10.1145/3430030.

[111] I. Novikau, E. A. Startsev, and I. Y. Dodin, "Quantum signal processing for simulating cold plasma waves," *Phys. Rev. A*, vol. 105, no. 6, p. 062444, Jun. 2022, doi: 10.1103/PhysRevA.105.062444.

[112] B. M. Boghosian and W. Taylor, "Quantum lattice-gas model for the many-particle Schrödinger equation in d dimensions," *Phys. Rev. E*, vol. 57, no. 1, pp. 54–66, Jan. 1998, doi: 10.1103/PhysRevE.57.54.

[113] J. Yepez, "Lattice-gas quantum computation," *Int. J. Mod. Phys. C*, vol. 09, no. 08, pp.

1587–1596, Dec. 1998, doi: 10.1142/S0129183198001436.

[114] J. Yepez, "Quantum lattice-gas model for computational fluid dynamics," *Phys. Rev. E*, vol. 63, no. 4, p. 046702, Mar. 2001, doi: 10.1103/PhysRevE.63.046702.

[115] J. Yepez, "Quantum lattice-gas model for the Burgers equation," *Journal of Statistical Physics*, vol. 107, no. 1/2, pp. 203–224, 2002, doi: 10.1023/A:1014514805610.

[116] M. M. Micci and J. Yepez, "Measurement-based quantum lattice gas model of fluid dynamics in 2+1 dimensions," *Phys. Rev. E*, vol. 92, no. 3, p. 033302, Sep. 2015, doi: 10.1103/PhysRevE.92.033302.

[117] J. Yepez, "Type-II quantum computers," *Int. J. Mod. Phys. C*, vol. 12, no. 09, pp. 1273–1284, Nov. 2001, doi: 10.1142/S0129183101002668.

[118] F. Gaitan, "Finding flows of a Navier–Stokes fluid through quantum computing," *npj Quantum Inf*, vol. 6, no. 1, p. 61, Jul. 2020, doi: 10.1038/s41534-020-00291-0.

[119] F. Gaitan, "Finding solutions of the Navier-Stokes equations through quantum computing— Recent progress, a generalization, and next steps forward," *Adv Quantum Tech*, vol. 4, no. 10, p. 2100055, Oct. 2021, doi: 10.1002/qute.202100055.

[120] F. Oz, O. San, and K. Kara, "An efficient quantum partial differential equation solver with chebyshev points," *Sci Rep*, vol. 13, no. 1, p. 7767, May 2023, doi: 10.1038/s41598-023-34966-3.

[121] L. Budinski, "Quantum algorithm for the advection–diffusion equation simulated with the lattice Boltzmann method," *Quantum Inf Process*, vol. 20, no. 2, p. 57, Feb. 2021, doi: 10.1007/s11128-021-02996-3.

[122] A. Mezzacapo, M. Sanz, L. Lamata, I. L. Egusquiza, S. Succi, and E. Solano, "Quantum simulator for transport phenomena in fluid flows," *Sci Rep*, vol. 5, no. 1, p. 13153, Aug. 2015, doi: 10.1038/srep13153.

[123] A. Engel, G. Smith, and S. E. Parker, "Linear embedding of nonlinear dynamical systems and prospects for efficient quantum algorithms," *Physics of Plasmas*, vol. 28, no. 6, p. 062305, Jun. 2021, doi: 10.1063/5.0040313.

[124] E. Farhi, J. Goldstone, and S. Gutmann, "A quantum approximate optimization algorithm," 2014, *arXiv*. doi: 10.48550/ARXIV.1411.4028.

[125] C. Bravo-Prieto, R. LaRose, M. Cerezo, Y. Subasi, L. Cincio, and P. J. Coles, "Variational quantum linear solver," 2019, *arXiv*. doi: 10.48550/ARXIV.1909.05820.

[126] P. J. Ollitrault *et al.*, "Quantum equation of motion for computing molecular excitation energies on a noisy quantum processor," *Phys. Rev. Research*, vol. 2, no. 4, p. 043140, Oct. 2020, doi: 10.1103/PhysRevResearch.2.043140.

[127] M. Cerezo *et al.*, "Variational quantum algorithms," *Nat Rev Phys*, vol. 3, no. 9, pp. 625–644, Aug. 2021, doi: 10.1038/s42254-021-00348-9.

[128] G. E. Crooks, "Gradients of parameterized quantum gates using the parameter-shift rule and gate decomposition," 2019, *arXiv*. doi: 10.48550/ARXIV.1905.13311.

[129] K. Mitarai, M. Negoro, M. Kitagawa, and K. Fujii, "Quantum circuit learning," *Phys. Rev. A*, vol. 98, no. 3, p. 032309, Sep. 2018, doi: 10.1103/PhysRevA.98.032309.

[130] M. Schuld, V. Bergholm, C. Gogolin, J. Izaac, and N. Killoran, "Evaluating analytic gradients on quantum hardware," *Phys. Rev. A*, vol. 99, no. 3, p. 032331, Mar. 2019, doi: 10.1103/PhysRevA.99.032331.

[131] D. Wierichs, J. Izaac, C. Wang, and C. Y.-Y. Lin, "General parameter-shift rules for quantum

gradients," *Quantum*, vol. 6, p. 677, Mar. 2022, doi: 10.22331/q-2022-03-30-677.

[132] M. Broughton *et al.*, "TensorFlow quantum: A software framework for quantum machine learning," 2020, *arXiv*. doi: 10.48550/ARXIV.2003.02989.

[133] S. Khatri, R. LaRose, A. Poremba, L. Cincio, A. T. Sornborger, and P. J. Coles, "Quantum-assisted quantum compiling," *Quantum*, vol. 3, p. 140, May 2019, doi: 10.22331/q-2019-05-13-140.

[134] J. R. McClean, J. Romero, R. Babbush, and A. Aspuru-Guzik, "The theory of variational hybrid quantum-classical algorithms," *New J. Phys.*, vol. 18, no. 2, p. 023023, Feb. 2016, doi: 10.1088/1367-2630/18/2/023023.

[135] J. A. Nelder and R. Mead, "A simplex method for function minimization," *The Computer Journal*, vol. 7, no. 4, pp. 308–313, Jan. 1965, doi: 10.1093/comjnl/7.4.308.

[136] A. Peruzzo *et al.*, "A variational eigenvalue solver on a photonic quantum processor," *Nat Commun*, vol. 5, no. 1, p. 4213, Jul. 2014, doi: 10.1038/ncomms5213.

[137] S. N. Genin, I. G. Ryabinkin, and A. F. Izmaylov, "Quantum chemistry on quantum annealers," 2019, *arXiv*. doi: 10.48550/ARXIV.1901.04715.

[138] S. Gocho *et al.*, "Excited state calculations using variational quantum eigensolver with spin-restricted ansätze and automatically-adjusted constraints," *npj Comput Mater*, vol. 9, no. 1, p. 13, Jan. 2023, doi: 10.1038/s41524-023-00965-1.

[139] Y. Y. Liu *et al.*, "Application of a variational hybrid quantum-classical algorithm to heat conduction equation and analysis of time complexity," *Physics of Fluids*, vol. 34, no. 11, p. 117121, Nov. 2022, doi: 10.1063/5.0121778.

[140] Y. Liu *et al.*, "A variational quantum algorithm-based numerical method for solving potential and Stokes flows," 2023, *arXiv*. doi: 10.48550/ARXIV.2303.01805.

[141] H.-L. Liu *et al.*, "Variational quantum algorithm for the Poisson equation," *Phys. Rev. A*, vol. 104, no. 2, p. 022418, Aug. 2021, doi: 10.1103/PhysRevA.104.022418.

[142] C. Cîrstoiu, Z. Holmes, J. Iosue, L. Cincio, P. J. Coles, and A. Sornborger, "Variational fast forwarding for quantum simulation beyond the coherence time," *npj Quantum Inf*, vol. 6, no. 1, p. 82, Sep. 2020, doi: 10.1038/s41534-020-00302-0.

[143] S. Chakrabarti, R. Krishnakumar, G. Mazzola, N. Stamatopoulos, S. Woerner, and W. J. Zeng, "A threshold for quantum advantage in derivative pricing," *Quantum*, vol. 5, p. 463, Jun. 2021, doi: 10.22331/q-2021-06-01-463.

[144] F. Fontanela, A. Jacquier, and M. Oumgari, "Short Communication: A quantum algorithm for linear PDEs arising in finance," *SIAM J. Finan. Math.*, vol. 12, no. 4, pp. SC98–SC114, Jan. 2021, doi: 10.1137/21M1397878.

[145] D. Herman *et al.*, "Quantum computing for finance," *Nat Rev Phys*, vol. 5, no. 8, pp. 450–465, Jul. 2023, doi: 10.1038/s42254-023-00603-1.

[146] P. Mocz and A. Szasz, "Toward cosmological simulations of dark matter on quantum computers," *ApJ*, vol. 910, no. 1, p. 29, Mar. 2021, doi: 10.3847/1538-4357/abe6ac.

[147] M. Lubasch, J. Joo, P. Moinier, M. Kiffner, and D. Jaksch, "Variational quantum algorithms for nonlinear problems," *Phys. Rev. A*, vol. 101, no. 1, p. 010301, Jan. 2020, doi: 10.1103/PhysRevA.101.010301.

[148] O. Kyriienko, A. E. Paine, and V. E. Elfving, "Solving nonlinear differential equations with differentiable quantum circuits," *Phys. Rev. A*, vol. 103, no. 5, p. 052416, May 2021, doi: 10.1103/PhysRevA.103.052416.

[149] A. C. Vazquez, "Extrapolation methods in Quantum Computing," ETH Zurich, 2022. doi: 10.3929/ETHZ-B-000586831.

# Appendix

## *A1*

```
from itertools import product
import numpy as np
from qiskit.quantum_info import Operator
from qiskit.circuit import library
import matplotlib.pyplot as plt

def pad_matrix(matrix):
    # Pad with 0 to make square matrix
    max_shape = max(matrix.shape[0], matrix.shape[1])
    deficiency = int(np.power(2,np.ceil(np.log2(max_shape)))) - max_shape
    if matrix.shape[0] != matrix.shape[1]:
        if matrix.shape[0] > matrix.shape[1]:
            pad_width = [(0, 0), (0, matrix.shape[0] - matrix.shape[1])]
        else:
            pad_width = [(0, matrix.shape[1] - matrix.shape[0]), (0, 0)]
        matrix = np.pad(matrix, pad_width)
    matrix = np.pad(matrix,[(0,deficiency),(0,deficiency)])
    return matrix

def decompose_pauli(matrix):
    matrix = pad_matrix(matrix)
    matrix_len = matrix.shape[0]
    nqubits = int(np.log2(matrix_len))

    pauli = {
        'x': Operator(library.XGate().to_matrix()),
        'y': Operator(library.YGate().to_matrix()),
        'z': Operator(library.ZGate().to_matrix()),
        'i': Operator(library.IGate().to_matrix())
    }

    decomposition = {}
    for permutation in product(*[list(pauli.keys())]*nqubits):
        permutation = "".join(permutation)
        base_matrix = pauli[permutation[0]]
        for idx in range(1, len(permutation)):
            base_matrix = base_matrix.tensor(pauli[permutation[idx]])

        decomposition_component = np.trace(np.dot(base_matrix, matrix)) / matrix_len
        if 0!=decomposition_component:
            decomposition[permutation] = decomposition_component

    return decomposition

max_n = 8
N = [2**n for n in range(1,max_n)]
sparsity = [len(decompose_pauli(2*np.eye(2**n)
```

```
                                    - np.diag(np.ones(2**n-1),-1)
                                    -  np.diag(np.ones(2**n-1),1)))   for   n   in
range(1,max_n)]
plt.plot(N,sparsity)
plt.xlabel('N')
plt.ylabel('Pauli Terms')
plt.title('Sparsity in Pauli Basis of NxN Laplacian')
plt.show()
```

## *A2*

```python
#!/usr/bin/python3

import qiskit
from qiskit_aer import Aer
from scipy.linalg import expm
import numpy as np
import matplotlib.pyplot as plt

min_t = 1
max_t = 5

# Number of Trotter Steps
m = [10,20,40,80,100, 200, 400, 800, 1000, 2000, 4000, 8000, 10000]

# 1st order Trotter
for t in range(min_t,max_t+1):
    # Calculate exact solution classically
    exact_solution = expm( -t * 1j * ( np.array([[0, 1], [1, 0]])
      + np.array([[1, 0], [0, -1]]) ) )

    errors = []
    for r in m:
        # Create a register of 1 qubit
        myQRegister = qiskit.QuantumRegister(1, '\psi')

        # Create a quantum circuit with using myRegister
        myCircuit = qiskit.QuantumCircuit(myQRegister)
        for _r in range(r):
            myCircuit.rx(t*2/r,0)
            myCircuit.rz(t*2/r,0)

        # Simulate the circuit to obtain overall unitary of Trotterization
        mySimulator = Aer.get_backend('unitary_simulator')
        result = mySimulator.run(myCircuit).result()
        finalUnitary = result.get_unitary()

        # Compare circuit unitary with exact unitary
        errors.append( np.linalg.norm(finalUnitary - exact_solution,2) )

    plt.loglog(m,errors)

# 2nd order Trotter
```

```python
for t in range(min_t,max_t+1):
    # Calculate exact solution classically
    exact_solution = expm( -t * 1j * ( np.array([[0, 1], [1, 0]])
      + np.array([[1, 0], [0, -1]]) ) )

    errors = []
    for r in m:
        # Create a register of 1 qubit
        myQRegister = qiskit.QuantumRegister(1, '\psi')

        # Create a quantum circuit with using myRegister
        myCircuit = qiskit.QuantumCircuit(myQRegister)
        for _r in range(r):
            myCircuit.rx(t/r,0)
            myCircuit.rz(t*2/r,0)
            myCircuit.rx(t/r,0)

        # Simulate the circuit to obtain overall unitary of Trotterization
        mySimulator = Aer.get_backend('unitary_simulator')
        result = mySimulator.run(myCircuit).result()
        finalUnitary = result.get_unitary()

        # Compare circuit unitary with exact unitary
        errors.append( np.linalg.norm(finalUnitary - exact_solution,2) )

    plt.loglog(m,errors,linestyle='dashed')

plt.xlabel('# of Trotter steps')
plt.ylabel('|error|_2')
plt.legend(['t = {}, 1st Order'.format(t) for t in range(min_t,max_t+1)]
      +['t = {}, 2nd Order'.format(t) for t in range(min_t,max_t+1)])
```

## A3

```python
import qiskit
from qiskit import QuantumCircuit, QuantumRegister, ClassicalRegister
from qiskit_aer import Aer
from qiskit.quantum_info.operators import Operator
import numpy as np
from scipy.linalg import fractional_matrix_power
from scipy.io import loadmat
from copy import deepcopy
from matplotlib import pyplot as plt

# Define the linear system:
# As an example, solve a matrix with numbers +,- 1...0.05  on the anti-diagonal,
# which is a non-Hermitian matrix
step = 0.05
A = np.diag(np.concatenate( (np.arange(-
```

```
1,0,step),np.arange(step,1+step,step)) ) )
# Take anti-transpose of A to demonstrate the non-Hermitian matrix case.
A = np.flipud(A)
# Save matrix to check againt classical solution later
A_orig = deepcopy(A)
# Normalize A. If an upper bound is known, use that instead.
# A = A/np.linalg.norm(A)

b = np.ones(np.shape(A)[1])
# Turn b into a quantum state
b = b/np.linalg.norm(b,2)
b_orig = deepcopy(b)

# Hermitian Dilation: only if A is not Hermitian
if np.any(A != A.conj().T):
    A = np.block([
        [np.zeros(np.shape(A)),A],
        [A.conj().T,np.zeros(np.shape(A))]
    ])
    b = np.block([
        b,
        np.zeros(np.shape(b))
    ])
    HD = True
else:
    HD = False

# The matrix A needs to padded to some 2^n to enable block-encoding
if np.size(A)>1:
    A_num_qubits = int(np.ceil(np.log2(np.shape(A)[0])))
    padding_size = 2**A_num_qubits - np.shape(A)[0]
    if padding_size > 0:
        A = np.block([
            [A, np.zeros([np.shape(A)[0],padding_size])],
            [np.zeros([padding_size,np.shape(A)[0]]),
np.zeros([padding_size,padding_size])]
        ])
else:
    A_num_qubits = 1
    padding_size = 1
    A = np.array([[A,0],[0,0]])
# Similarly, pad b
b = np.pad(b,(0,padding_size))

# Define the block-encoding of the matrix A
# If you have an efficient circuit to realize O_A (or U_A), use it here
O = np.block([
    [A    ,  -fractional_matrix_power(np.eye(np.shape(A)[0]) -
np.linalg.matrix_power(A,2),0.5)],
    [fractional_matrix_power(np.eye(np.shape(A)[0]) -
np.linalg.matrix_power(A,2),0.5)    ,    A]
])
# We also need to get the block-encoding size, i.e. m, used to encode A in
```

```
U_A
m = int(np.log2(np.shape(O)[0]) - A_num_qubits)
O_A_num_qubits = int(np.log2(np.shape(O)[0]))
# Create the operator O_A in Qiskit
operatorA = Operator(O)

# Create the three registers for QSP:
# 1) 1 Z rotation qubit
# 2) m block-encoding ancillas
# 3) register for b
register_1 = QuantumRegister(size = 1, name = '|0>')
register_2 = QuantumRegister(size = m, name = '|0^m>')
register_3 = QuantumRegister(size = O_A_num_qubits-m, name = '|\phi>')

# Create a rotation circuit in the block-encoding basis
def CR_phi_d(phi, d, register_1, register_2):
    circuit = QuantumCircuit(register_1,register_2,name = 'CR_( \phi \tilde
{})'.format(d))
    circuit.cx(register_2,register_1,ctrl_state=0)
    circuit.rz(phi*2, register_1)
    circuit.cx(register_2,register_1,ctrl_state=0)
    return circuit

# Load QSP angles
phi_angles = np.array( loadmat('phi_k_50.mat') ).item()['phi']
phase_angles = phi_angles.reshape(phi_angles.shape[0])

# Create QSP circuit
QSP_circuit = QuantumCircuit(register_1, register_2, register_3, name =
'QSP')
# Initialize state |b>. If you have an efficient implementation for b, it
goes here
QSP_circuit.initialize(b,list(reversed(register_3)))

# First Hadamard the ancilla qubit since we want Re(P(A))
QSP_circuit.h(register_1)
# Note: QSPPACK produces symmetric phase angles, so reversing phase angles is
unnecessary
for d, phi in reversed(list(enumerate(phase_angles))):
    QSP_circuit = QSP_circuit.compose(CR_phi_d(phi,d,register_1,register_2))
    if d>0:
        # The endianness of the bits matters. Need to change the order of the
bits
        QSP_circuit.append(operatorA,list(reversed(register_3[:])) +
```

```
register_2[:])
# Apply the final Hadamard gate
QSP_circuit.h(register_1)
# Account for little vs. big endian
QSP_circuit = QSP_circuit.reverse_bits()

# Run statevector simulator
solver='statevector'
backend = Aer.get_backend('statevector_simulator',precision = "double")
job = backend.run(QSP_circuit, shots=0)

# Extract relevant portion of statevector
QSP_statevector = job.result().get_statevector()
if HD:
    P_A_b =
np.real(QSP_statevector.data[int(b_orig.shape[0]):(2*b_orig.shape[0])])
else:
    P_A_b = np.real(QSP_statevector.data[0:b.shape[0]])
P_A_b = P_A_b/np.linalg.norm(P_A_b)

# Get expected result using classical solver
expected_P_A_b = np.linalg.solve(A_orig,b_orig)
expected_P_A_b = expected_P_A_b/np.linalg.norm(expected_P_A_b)

# Plot QSP polynomial
x = np.flipud(A_orig).diagonal()
fig, ax1 = plt.subplots()
ax1.set_title('QSP QLSA')

ax1.scatter(x,P_A_b/P_A_b[-1],marker='x',c='g')
ax1.scatter(x,expected_P_A_b/expected_P_A_b[-1],marker='o',facecolors='none',
edgecolors='k')
ax1.set_ylabel('P(x), 1/x')
plt.legend(['P(x)','1/x'],loc = 2)

ax2 = ax1.twinx()
ax2.plot(x[:x.size//2],np.log10(np.abs((P_A_b[:x.size//2]-
expected_P_A_b[:x.size//2])/expected_P_A_b[-1])),'r')
ax2.plot(x[x.size//2:],np.log10(np.abs((P_A_b[x.size//2:]-
expected_P_A_b[x.size//2:])/expected_P_A_b[-1])),'r')
ax2.set_ylim(bottom=-12, top=0)
ax2.set_ylabel('log10 |P(x)-1/x|')
plt.legend(['error'],loc = 1)

plt.xlabel('x')
plt.show()
```